\documentclass[aps,pre,amsmath,twocolumn,floatfix]{revtex4}
\usepackage{graphicx} 
\usepackage{natbib}

\begin{document}
\newcommand{\btau}{\mbox{\boldmath{$\tau$}}}

\title{Anisotropy driven dynamics in vibrated granular rods}

\author{Dmitri Volfson$^1$, Arshad Kudrolli$^2$ and Lev S. Tsimring$^1$}
\affiliation{$^1$ Institute for Nonlinear Science, University of California,  
San Diego, La Jolla, CA 92093-0402 \\ $^2$ Department of Physics, Clark University, MA 01610}
\date{\today}

\begin{abstract} 
The dynamics of a set of rods bouncing on a vertically vibrated plate is
investigated using experiments, simulations, and theoretical analysis.
The experiments and simulations are performed within an annulus
to impose periodic boundary conditions. Rods tilted with
respect to the vertical are observed to spontaneously develop a
horizontal velocity depending on the acceleration of the plate. For high
plate acceleration, the rods are observed to always move in the
direction of tilt. However, the rods are also observed to move 
opposite to direction of tilt for a small range of plate acceleration
and rod tilt. A phase diagram of the observed motion is presented as a
function of plate acceleration and the tilt of the rods which is varied
by changing the number of rods inside the annulus. Next we introduce a
novel molecular dynamics method to simulate the dynamics of the rods 
using the dimensions and dissipation parameters from the experiments. We reproduce the
observed horizontal rod speeds as a function of rod tilt and plate
acceleration in the simulations. By decreasing the friction between the
rods and the base plate to zero in the simulation, we identify the
friction during the collision as the crucial ingredient for occurrence
of the horizontal motion. Guided by the data from the experiments and
the simulations, we construct a mechanical model for the dynamics of the
rods in the limit of thin rods. The starting point of the analysis is
the collision of a single rod with an oscillating plate. Three friction
regimes are identified: slide, slip-stick, and slip reversal. A formula
is derived for the observed horizontal velocity as a function of tilt
angle. Good agreement for the horizontal velocity as a function
of rod tilt and plate acceleration is found between
experiments, simulations and theory.
\end{abstract} 

\maketitle

\section{Introduction}

Granular materials come in all shapes and sizes. Idealized spherical
particles, have been typically used to unravel the fascinating
properties displayed by granular materials. In vibrated granular
systems, periodic pattern formation, cluster formation, and complex size
separation have been observed. However, anisotropic grains are nearly as
numerous, and experience with thermal systems teaches us that shape
matters.  Only a handful of investigations have studied the impact of
anisotropy on granular systems, but the ones that have been accomplished
point to a rich phenomenology. 

For example, theoretical and numerical study of a low-density gas of
hard inelastic needles~\cite{huthmann99} shows two distinct regimes of
cooling related to the different scaling of rotational and translational
energy dissipation. In compaction experiments with granular
rods~\cite{villaruel00} granular rods vibrated in a tall narrow
container were observed to form ordered stacks i.e. a smectic phase
similar to that found in thermal systems, with the additional novelty
that the rods align with the gravitation field.  More recently,
self-organization of vortices was observed to occur when a shallow bed
of granular rods was vibrated \cite{blair03}. It was further shown, that the tilt of
the rod and vertical vibration was important to the occurrence of the
novel dynamics. While a phenomenological model of formation and growth
of the vortices has been proposed \cite{aranson03}, a detailed understanding of why the
rods move horizontally on a vertically vibrated plate was not reached. 

The collective motion of vibrated anisotropic grains is of considerable
interest as an example of spontaneous ratchet formation 
in a non-equilibrium dissipative system. Transport of thermal
particles in systems with microscopically asymmetric potential has been
studied in a number of recent publications~\cite{rachets}.  In
Ref.~\cite{farkas99}, ratchet transport was demonstrated for spherical
grains on a vertically vibrated asymmetric saw-tooth profile. In
Ref.~\cite{wambaugh02}, the transport of elongated grains on a
vertically vibrated ratchet-shaped plate has been studied numerically.
However, as follows from results of~\cite{blair03,aranson03}, and further described
in this paper, the transport of anisotropic grains may occur even
without externally imposed microscopic asymmetry of forcing, in which case 
the direction of motion is chosen as a result of spontaneous symmetry breaking.

In this paper, we apply experimental and numerical tools to a 
system of rods in a vibrated annular container to elucidate 
the development of coherent horizontal dynamics in anisotropic systems.
This geometry was specifically chosen to simplify the phenomenology in
order to focus on the mechanisms for the observed dynamics. Our
theoretical model is developed for even simpler quasi-two-dimensional
geometry where all rods are confined to a vertical plane, and periodic
boundary conditions are imposed. The theory is based on a detailed
description of frictional collisions between rods and the
vibrating plate which makes use of the assumption of constant kinematic
restitution coefficient. While the issues of restitution coefficient in
application to frictional impact of asymmetric bodies are still debated
in the literature (see, for example, \cite{stronge}), we show that even
this simplest model agrees very well with soft particle molecular
dynamics simulations of individual collisions.  To describe the
collective motion of the rods, we take advantage of the observation that
in the regime of stationary translation the mean horizontal momentum transfer 
due to the collision with bottom plate is zero, and assume that this
condition holds for a typical collision. Our simulations show that
during the flight the angular momentum of a single rod is transfered to
other rods, so the angular velocity of the rod at the end of the flight
becomes small and can be neglected. Furthermore, based on our numerical
simulations we make the assumption that the vertical velocity of the
center of mass (CM) just before the collision is equal to the CM
velocity in the beginning of the flight. These assumptions allow us to
find the mean translation CM velocity in a closed form.  We show that
this theory captures the essential mechanisms of the transport of tilted
rods on a vibrating plate.  

The paper is organized as follows. First, we introduce the experimental
system and report the observed dynamics as a function of the control
parameters such as plate acceleration and tilt of rods. Next we discuss the
molecular dynamics simulations corresponding to experimental parameters
and compare the results with the experimental data. Then using the data
as a guide, we construct a theoretical model for the occurrence of
spontaneous horizontal motion, and its dependence on control parameters,
and compare the results with the data. Finally, we discuss the general
implications of this study.

\section{Experiments}

\begin{figure}\begin{center}
\includegraphics[width=2.5in]{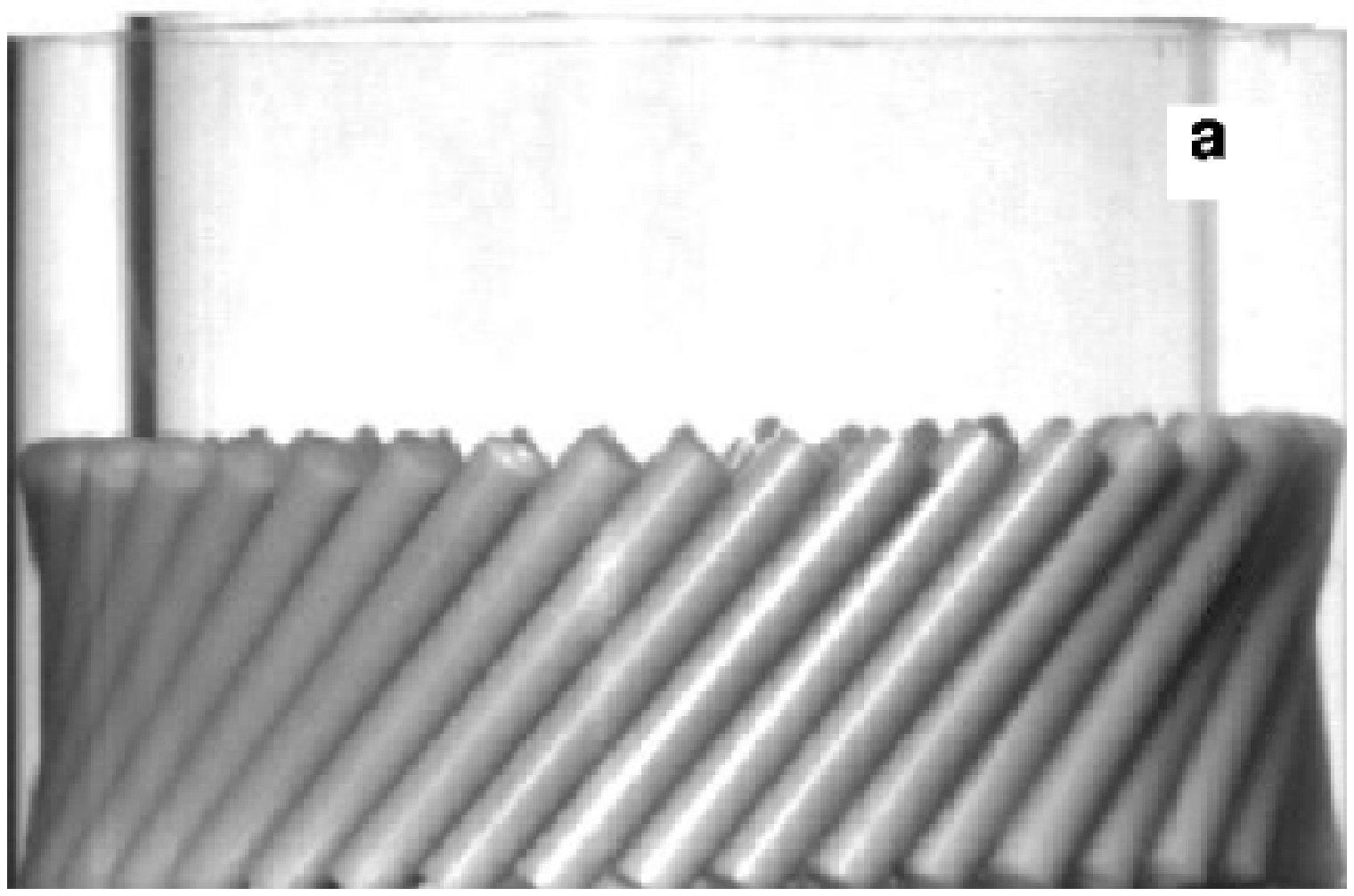}
\vskip 0.5cm
\includegraphics[angle=0,width=2.5in]{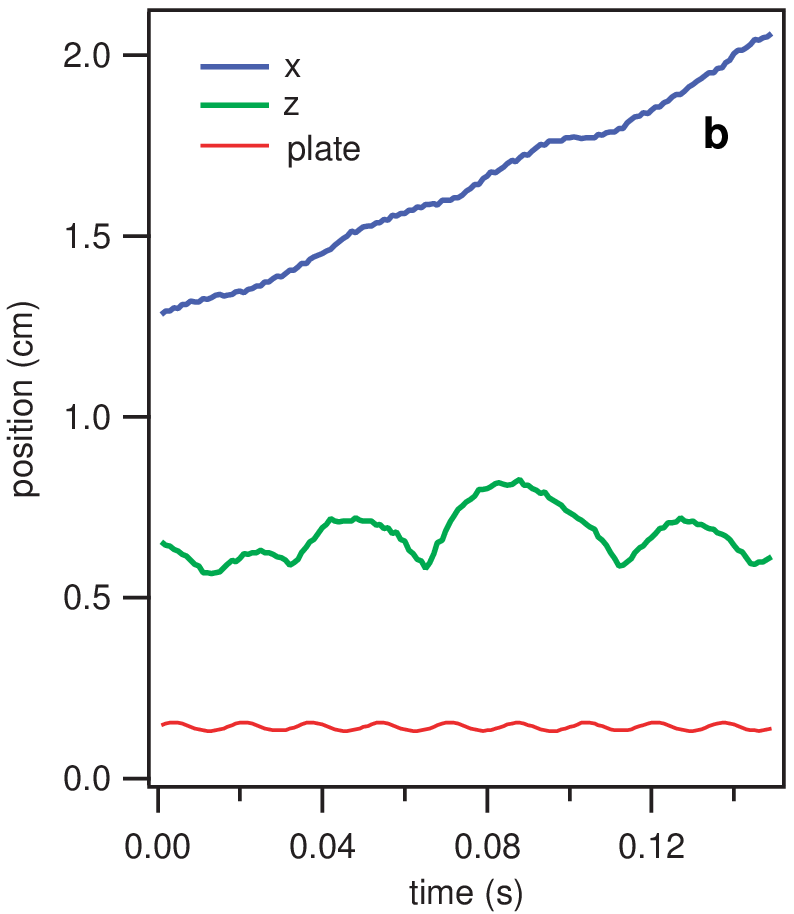}
\vskip 0.5cm
\caption{
(a) Annular geometry used in the experiments. 
(b) The horizontal and vertical position of
the rod end at the bottom as a function of time. The position of the
oscillating bottom plate is also plotted for reference. The three plots
are shifted for clarity. ($\Gamma = 3.3$, $N=50$, $\phi = 13^0$)}
\label{exper1}
\end{center}\end{figure}

The experimental system consists of an annular container with an inner
diameter of 7.28 cm and outer diameter of 9.45 cm, and is similar to
that used in Ref.~\cite{blair03}. An image is shown in
Fig.~\ref{exper1}(a). The sides are composed of clear acrylic and the
base plate is made of Aluminum. The container is attached to an
electromagnetic shaker through a linear bearing which allows only
vertical motion. A frequency generator along with a lock-in amplifier is
used to excite the system with a fixed frequency and amplitude. The data
reported here for frequency $f = 60$ Hz, and we note that qualitatively
similar behavior is observed when the frequency is varied between 50
and 100 Hz. The acceleration of the container is measured with an
accelerometer and is reported in terms of $\Gamma$, the measured peak
acceleration divided by the acceleration due to gravity. The tilt to the
rods is characterized by the angle $\phi$ with respect to the vertical
axis.

The rods used for the experiments are cylindrical with a
diameter of 0.635 cm and length 5.08 cm. One of the ends is
semi-spherical with the radius equal to the radius of the rod, and the
other end is flat. The rods are made of a Delrin and Teflon composite,
and the measured dissipation coefficients are as follows. The
coefficient of static and kinetic friction between the rod and the base
plate is 0.36 and 0.25 respectively. The coefficient of restitution is obtained 
by measuring the kinetic energy of the rod just before and after a collision with a stationary base 
plate. For nearly normal incidence ($\phi < 5^0$), the coefficient of 
restitution is approximately $0.8 \pm 0.1$ (see also~\cite{Blair04}.)  When a single rod
is placed inside the annulus, the maximum tilt angle 
that it can have is 53$^0$ due the curvature of the
annulus. By increasing the total number of rods $N$ in the annulus
from $N =$ 23 to 56, the tilt $\phi$ can be
varied from 53$^0$ to 0$^0$. 

Figure~\ref{exper1}(b) shows the typical motion of the bottom tip of a
rod as a function of time. The data is obtained by using a high-speed
Kodak digital camera with a frame rate of 1000 per second, and tracking the end of the
rod with appropriate use of lighting through the transparent side walls.
The vertical position of the tip $z$ is observed to oscillate, as the
rod bounces on the vibrating plate [also shown in the Fig.~\ref{exper1}(b)]. The
flight time is observed to vary and although the rod appears to almost always hit the
plate on its upstroke, a distribution of phases is observed. We
will discuss this issue further after introducing the molecular dynamics
simulations in a later section. On the other hand, the horizontal
position of the tip is observed to increase approximately linearly
although some oscillatory motion is also seen be superposed. The slope
of the $x$-position gives the average horizontal velocity which is
consistent with dividing the average circumference of the annulus by the
amount of time taken by the rods to go around once. This second method
is used to report observed average horizontal velocities. 

\begin{figure}\begin{center}
\includegraphics[angle=0,width=2.3in]{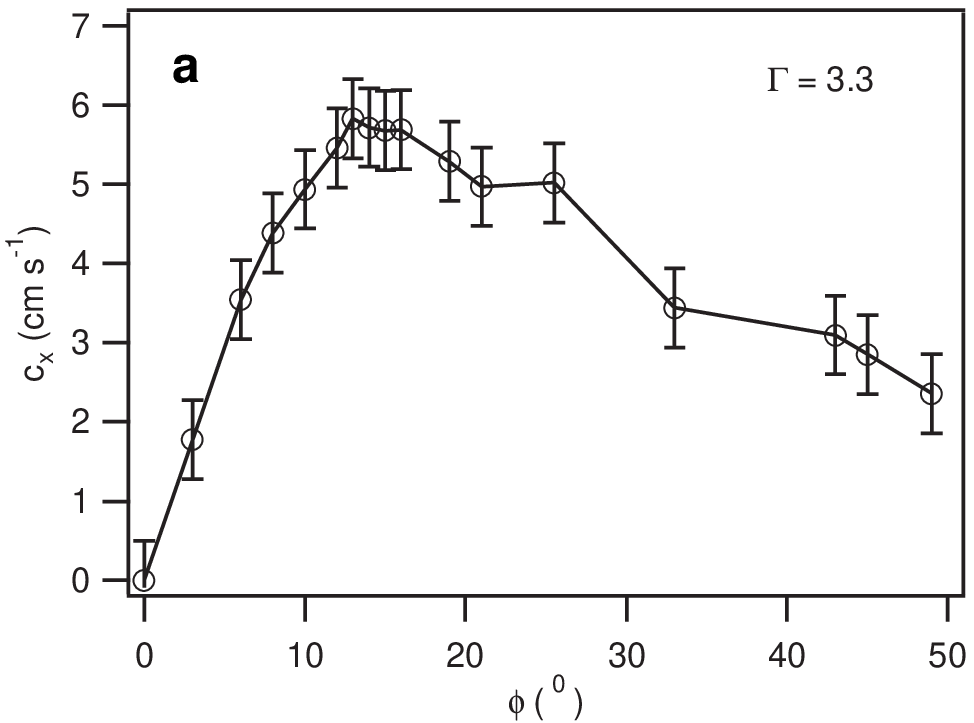}
\includegraphics[angle=0,width=2.5in]{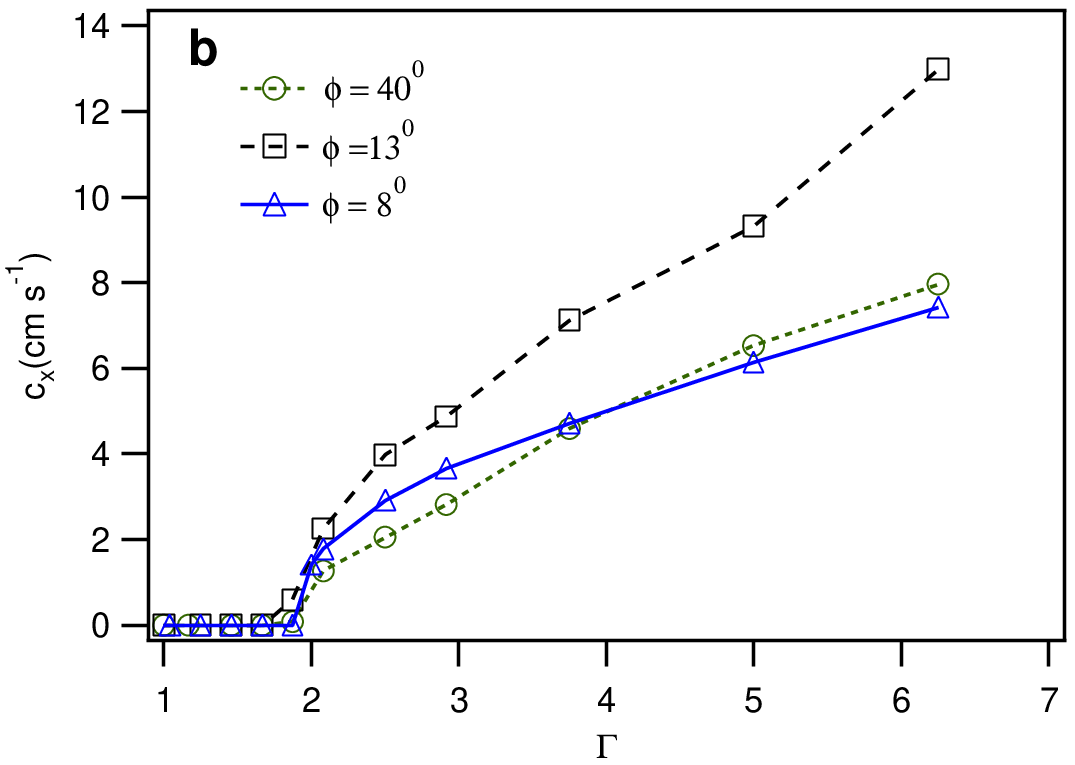}
\includegraphics[angle=0,width=2.6in]{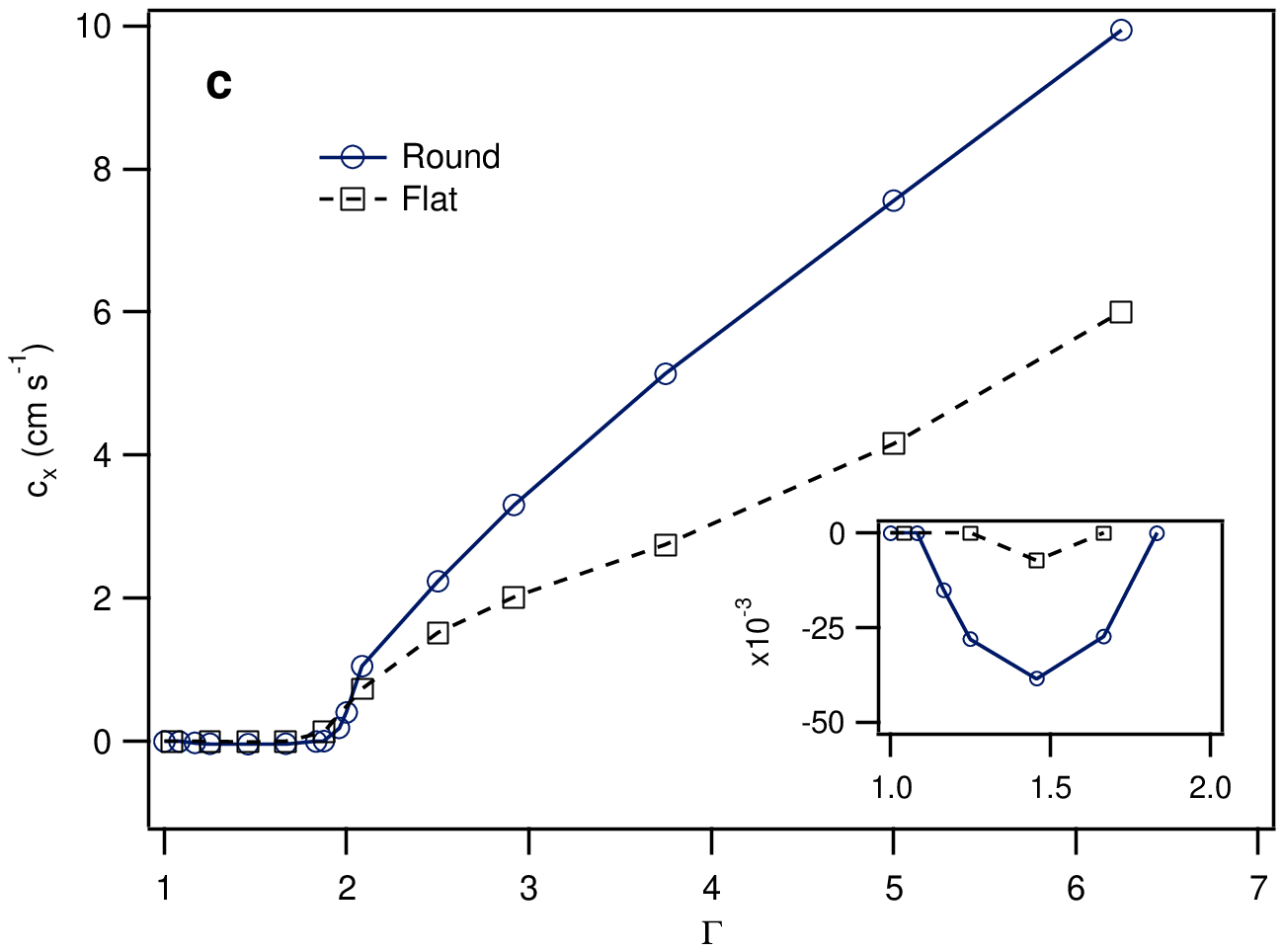}
\includegraphics[angle=0,width=2.5in]{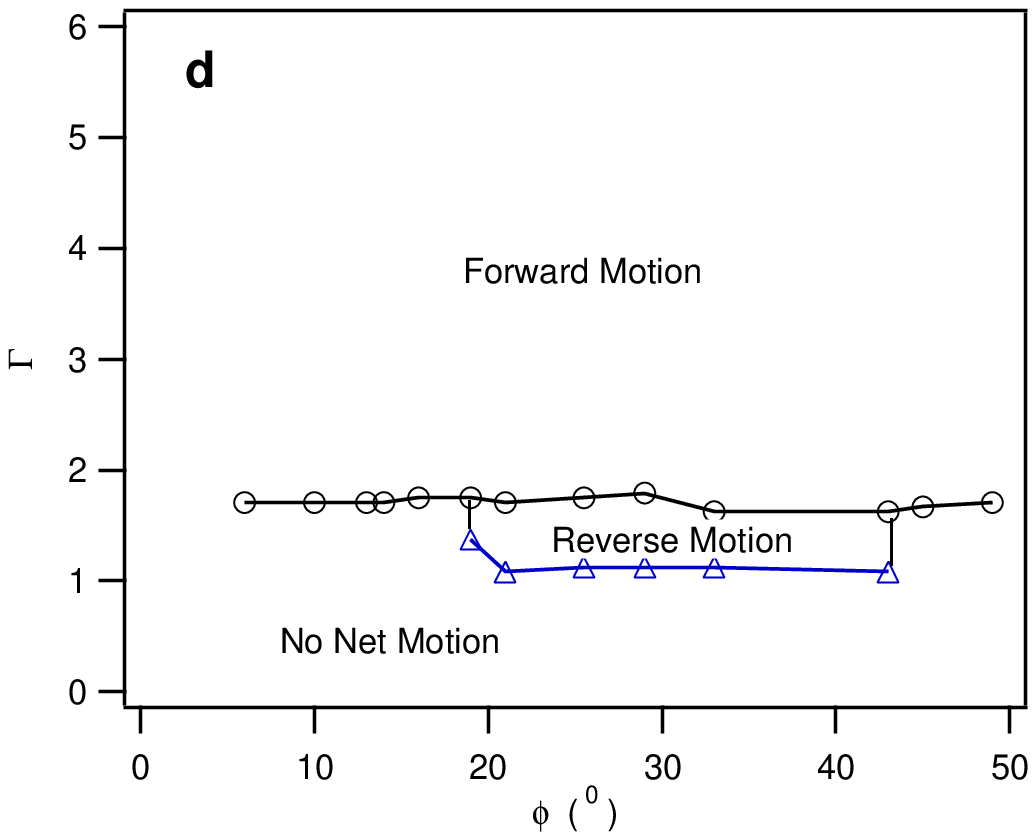}
\vskip 0.5cm
\caption{(a) The average horizontal velocity $c_x$ of the rods as a
function of tilt angle $\phi$. (b) $c_x$
versus $\Gamma$ for three different tilt angles. (c) The effect of
rod end shape on measured $c_x$. A flat end rod is observed to move slowly
compared to a rod with a round end. Horizontal motion in a direction opposite to the tilt is observed over a small range of low $\Gamma$. ($\phi = 28^0 \pm 2^0$) Inset: $c_x$ 
measured between $\Gamma$ = 1 and 2 is replotted to clarify the small magnitude
of the reverse motion.  (d) A phase diagram denoting the kinds of
horizontal velocity observed relative to the direction of the tilt. The
rods are observed to move in the direction of the tilt under most
conditions.}
\label{exper2}
\end{center}\end{figure}

The measured horizontal speed $c_x$ as a function of the tilt of the
rods $\phi$ for a fixed $\Gamma=3.3$ is shown in Fig.~\ref{exper2}(a).
$c_x$ is observed to increase from zero to a peak value and then
decrease. The rods are observed to always move in the same horizontal
direction as the tilt. It is to be noted that the error in determining $c_x$ arises from run to run variability due to slight differences in packing inside the annulus rather than from the actual measurement of the velocity itself. In this case, seven separate realizations were used to arrive at the average value of $c_x$ for a particular number of rods. 

By varying $\Gamma$ for a few values of $\phi$, we obtain its impact on observed horizontal velocities [see Fig.~\ref{exper2}(b).] Positive horizontal velocity is observed to
commence only above a finite $\Gamma \sim 1.6$ is reached. Below 
$\Gamma \sim 1.6$, no net horizontal velocity occurs except for a
narrow parameter range where a horizontal motion in a direction 
opposite to the tilt is observed. As shown in the inset to Fig.~\ref{exper2}(c),
this reverse horizontal motion is observed to be two orders of magnitude
slower than the forward motion. 

A phase diagram of the various kinds of observed motion is shown in Fig.~\ref{exper2}(d) as a
function of $\Gamma$ and $\phi$. Forward motion indicates parameters for which horizontal motion occurs in the direction of tilt, and reverse motion, indicates when the motion is in the opposite direction. The reverse motion is
observed to occur only for a narrow range of parameters. The horizontal
motion is predominately along the direction of tilt provided a minimum
acceleration for the container is achieved. 

To study the impact of the shape of the rod tip, we also performed
experiments by flipping the rods so that the flat end is at the bottom.
Figure~\ref{exper2}(c) compares the measured $c_x$ as a function of
$\Gamma$ for rods with rounded and flat ends but otherwise under
identical conditions. When the flat end interacts with the bottom plate,
the observed horizontal velocities are lower, the minimum rod tilt
required to obtain horizontal motion is $8^0$, but otherwise the
qualitative phenomena remains the same. 

\section{Numerical simulations}
We performed a series of numerical simulations of the bouncing rods on a
vibrated plate.  The rods are modeled as spherocylinders of
diameter $d$, length $l$, mass $m$, and moment of inertia $I$.
A rod has three translational and two rotational degrees of freedom, the
rotation of a rod around its own axis is neglected.  Our numerical
algorithm is based on the ``soft spheres'' molecular dynamics
technique~\cite{mdalgs}. The interaction forces between colliding
spherocylinders are calculated via the interaction between viscoelastic
virtual spheres of diameter $d$ centered at the closest points between
the axes of spherocylinders, so that the cylinders are in contact
whenever virtual spheres are. The normal forces between
virtual spheres are computed using Hertzian model and the tangential
frictional forces are computed by the Cundall-Struck algorithm. They
lead to the kinematic restitution coefficient of about 0.65-0.7 slightly
varying with impact angle.  In most cases we used equal friction
coefficients $\mu_{rr}=\mu_{rb}=\mu_{rw}=0.3$ for all sliding surfaces 
(rod-rod, rod-bottom and
rod-side wall) and ignored the difference between dynamic and static
friction coefficients while comparing to experiments.  The forces
arising from the interaction of virtual spheres are then applied to the
rods (see Appendix A for details).  The motion of rods was obtained by
integrating the Newton's equations with the forces and torques produced
by interactions of a rod with all the neighboring rods, walls of the
container, and by gravity.

\begin{figure}
\begin{center}
\includegraphics[width=3.0in]{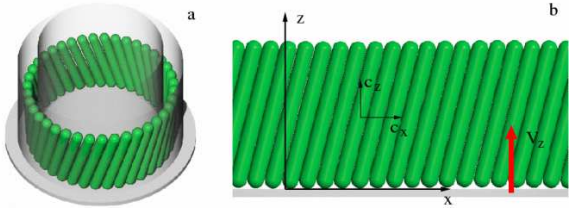}
\caption{Two types of geometry used in simulations: (a) - annulus geometry, (b) - 
quasi-2D geometry with periodic boundary conditions along $x$ where the motion of rods
is constrained to the $x-z$ plane.
}
\label{fig:num_setup}
\end{center}
\end{figure}

\begin{figure}
\begin{center}
\includegraphics[width=2.6in]{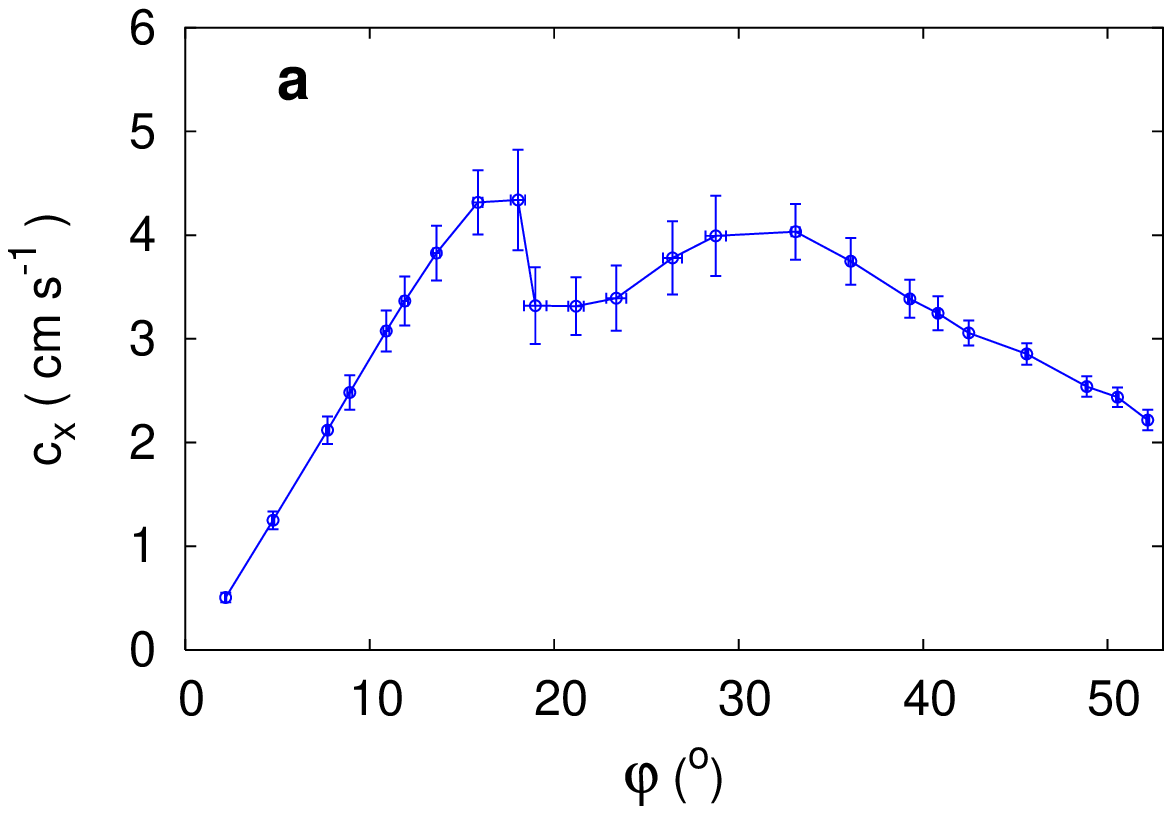}
\includegraphics[width=2.6in]{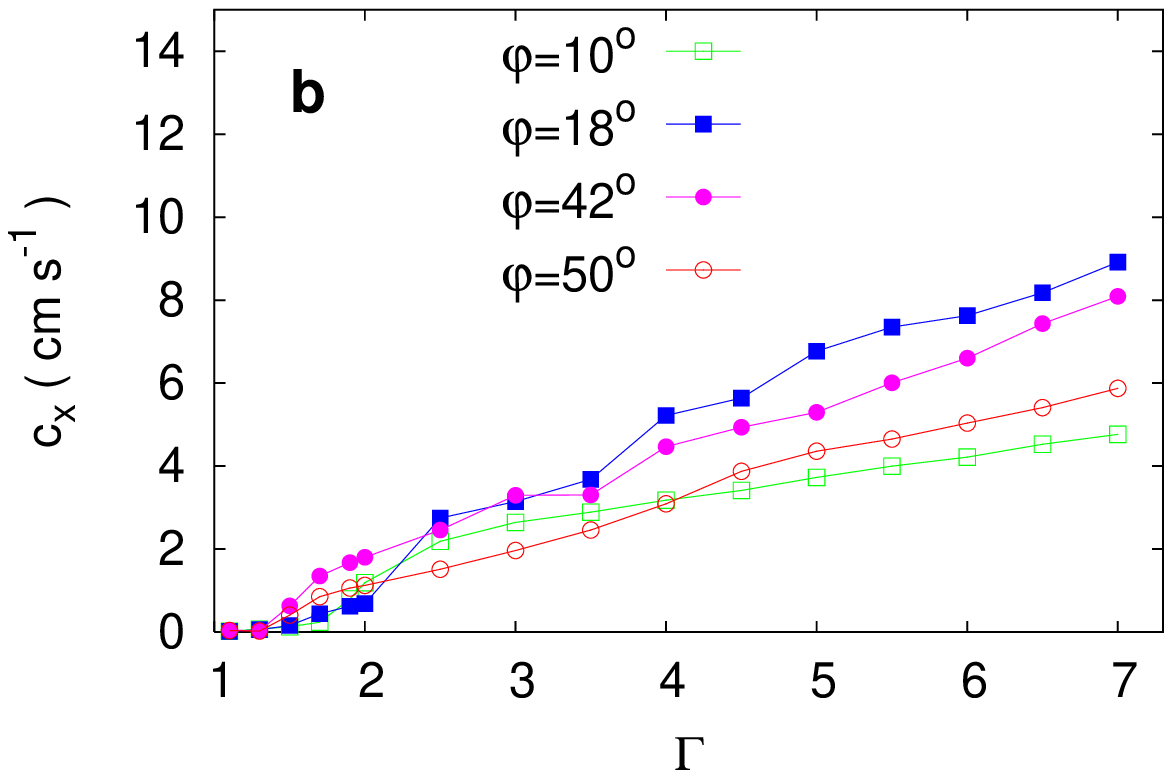} 
\includegraphics[width=3.0in]{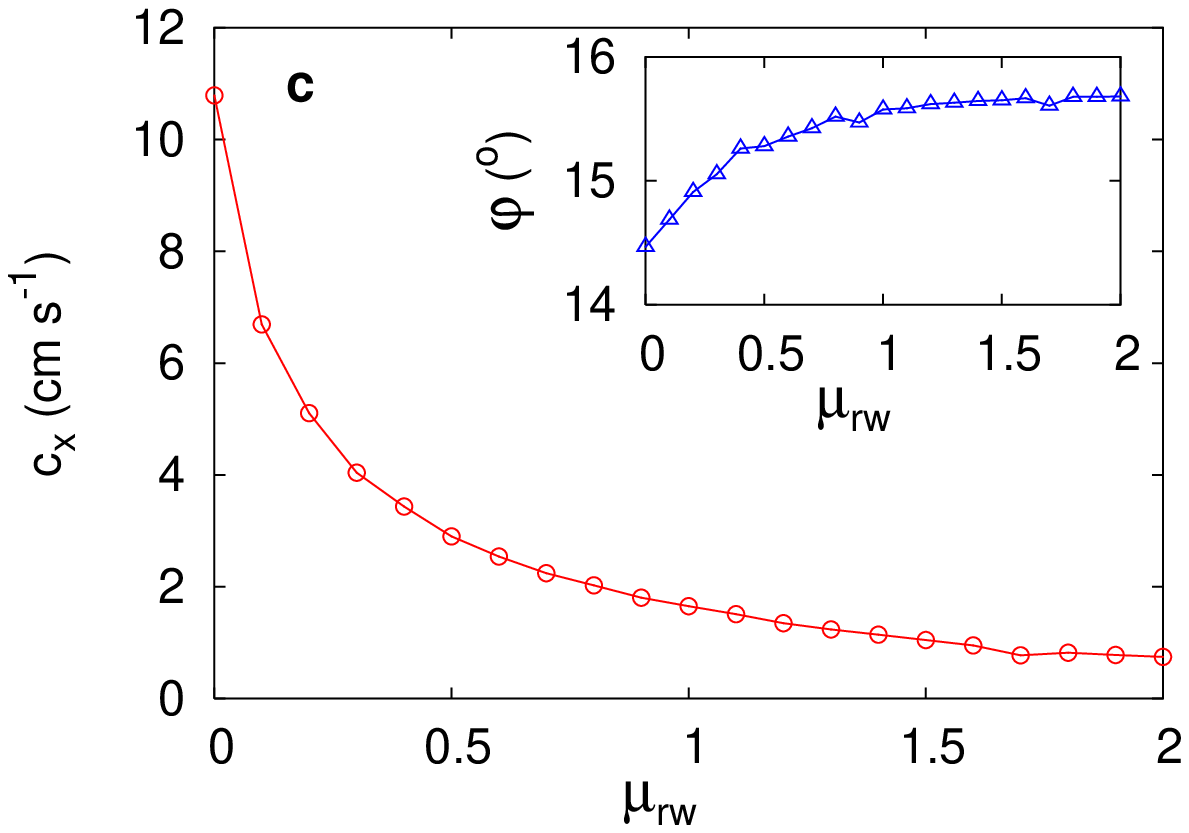} 
\includegraphics[width=3.0in]{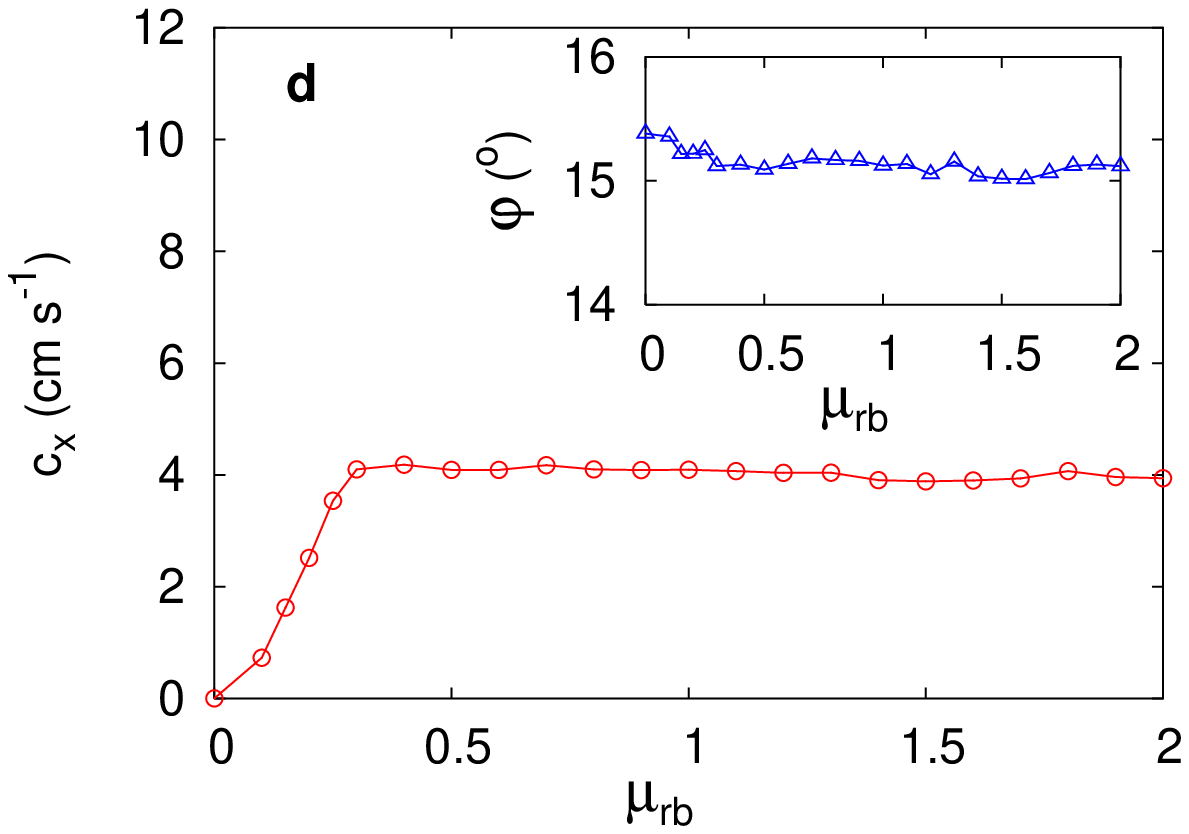} 
\caption{
The results of simulations in the annular geometry:
(a) - average translational velocity of rods as a function of tilt at $f=60Hz$,$\Gamma=3.3$; 
(b) - {\em idem} as a function of $\Gamma$ for a number of tilt angles;
(c) - {\em idem} and average tilt as a function of the coefficient of friction
between the rods and the walls; 
(d) - {\em idem} and average tilt as a function of the coefficient of friction
between the rods and the bottom.
}
\label{fig:md3d}
\end{center}
\end{figure}

\begin{figure}
\begin{center}
\includegraphics[width=1.3in]{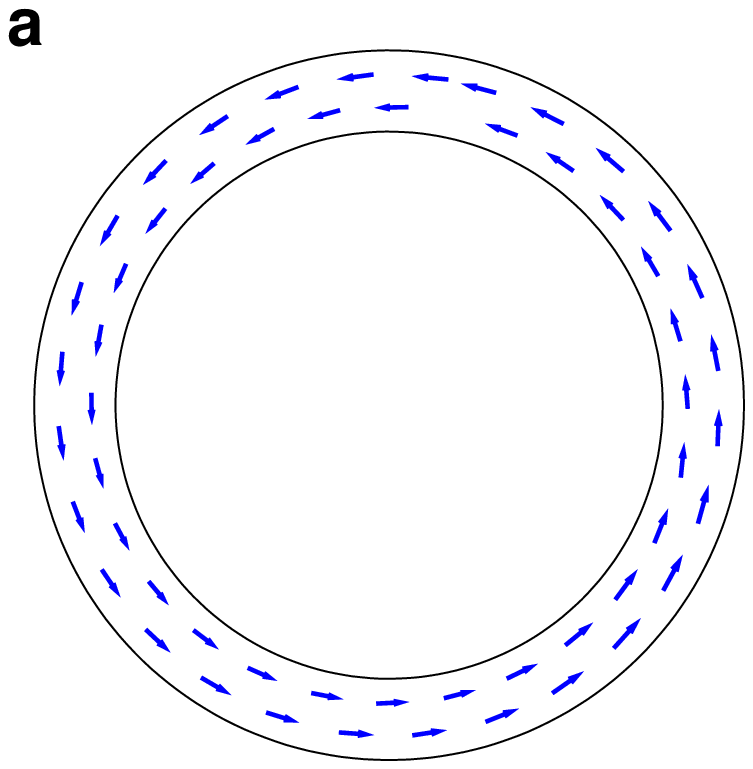}
\includegraphics[width=1.3in]{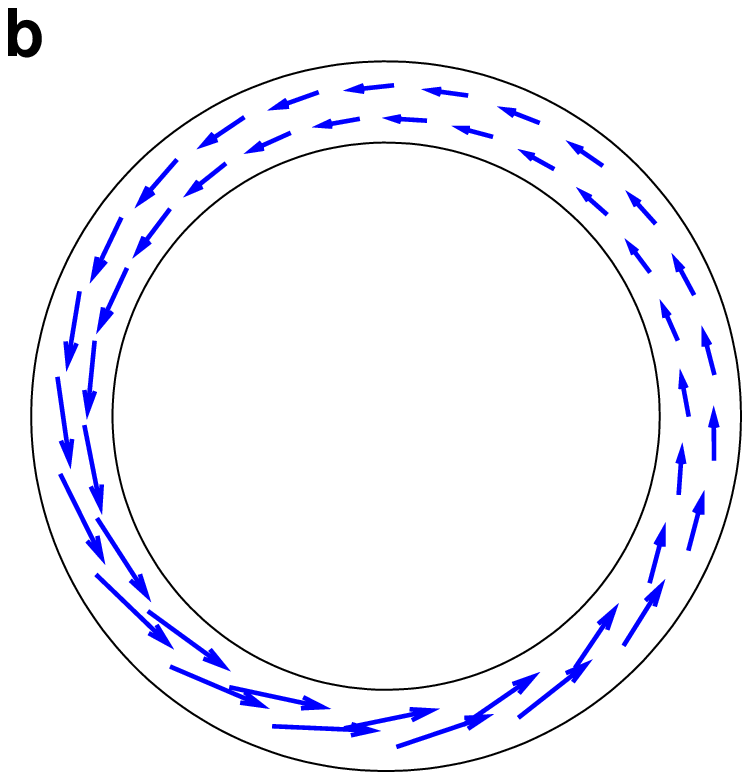}
\caption{
The results of simulations in the annular geometry illustrating formation of the
non-uniformity in the tilt for $f=60Hz$, $\Gamma=3.3$.
Shown are (normalized) horizontal projections of the rods' directors 
for two numbers of rods  (a) $N=55$, (b) $N=44$, the mean tilts are 
$\phi\approx9^o$ and $\phi\approx22^o$  respectively.
}
\label{fig:md3ddef}
\end{center}
\end{figure}

Figure~\ref{fig:num_setup} illustrates two configurations employed in
numerical simulations. An annular geometry [Fig.~\ref{fig:num_setup}(a)]
was used to match closely the experimental setup. However, to separate
the effects of side wall friction and oblique collisions among the rods
due to annulus curvature, we also studied the quasi-2D geometry of
Fig.~\ref{fig:num_setup}(b) in which axes of all rods are constrained to
the $x-z$ plane. This geometry is more amenable to the theoretical
analysis and was used for comparison with our theoretical predictions.

In simulations we observed robust drift of the rods in the direction of
inclination in agreement with experiments~\cite{movies}.
Figure~\ref{fig:md3d}(a) shows the translation velocity as a function of
the inclination angle for the annulus geometry and parameters $f=60$ Hz
and $\Gamma=3.3$ used in experiments. The translation velocity grows
linearly for small $\phi$, reaches maximum at $\phi\approx 18^o$ and
also for $\phi \approx 35^o$ after which  it decays to zero at large
$\phi$. For intermediate inclinations, $18^o < \phi < 35^o$, in most
cases we observed noticeable slowdown which however is not typically
observed in experiment. We will discuss the source of this discrepancy
below. 

Figure~\ref{fig:md3d}(b) shows the dependence of the average
translational velocity of rods on the acceleration of the container at
fixed frequency for a number of inclinations. As in experiments the
motion starts above a (slightly lower) threshold $\Gamma > 1.5$. Above
$\Gamma \approx 2.0$ this velocity grows roughly linearly with the
acceleration.  We explain the presence of the threshold by the friction
with side walls. Indeed, in the experiments in quasi-2D geometry  there
is no threshold and $c_x \propto \Gamma-1$ right down to $\Gamma=1.0$.
Figure \ref{fig:md3d}(c) depicts explicitly how the average
translational velocity  and average tilt depend on the coefficient of
friction with the sidewall $\mu_{rw}$, for a fixed number of rods. As
one could expect this dependence shows monotonic decay when $\mu_{rw}$
is increased; less expected is more than tenfold difference in
velocities for large and small $\mu_{rw}$ which probably explains
systematically slower translational velocities in simulations and
underlines an importance of the proper account of friction with walls.

Next, to eliminate the effects of the curved side walls and out-of-plane 
rod-rod collisions, we simulated the collective rod motion in a quasi-2D geometry with
periodic boundary conditions along the direction of the rod tilt [Figure
\ref{fig:num_setup}(b)].  All rods are confined strictly to the $x-z$
plane, and the interaction with side walls was ignored, while friction with the
vibrating bottom plate and among the rods was taken into account. We used a
fixed number of rods ($N=40$) and varied the length of the container, thereby
changing the tilt angle $\phi$ [see Figure \ref{fig:num_setup}(c)]. The relation between
the length and the tilt is well described by a simple formula $\cos\phi=dN/L$.

\begin{figure}[h]
\begin{center}
\includegraphics[angle=0,width=3.3in]{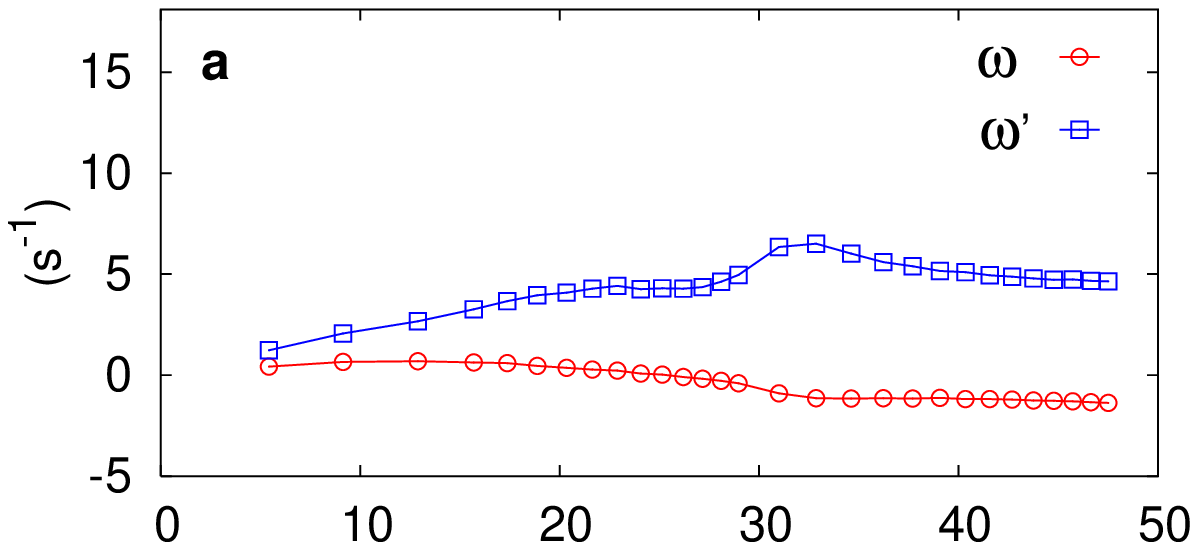}
\includegraphics[angle=0,width=3.3in]{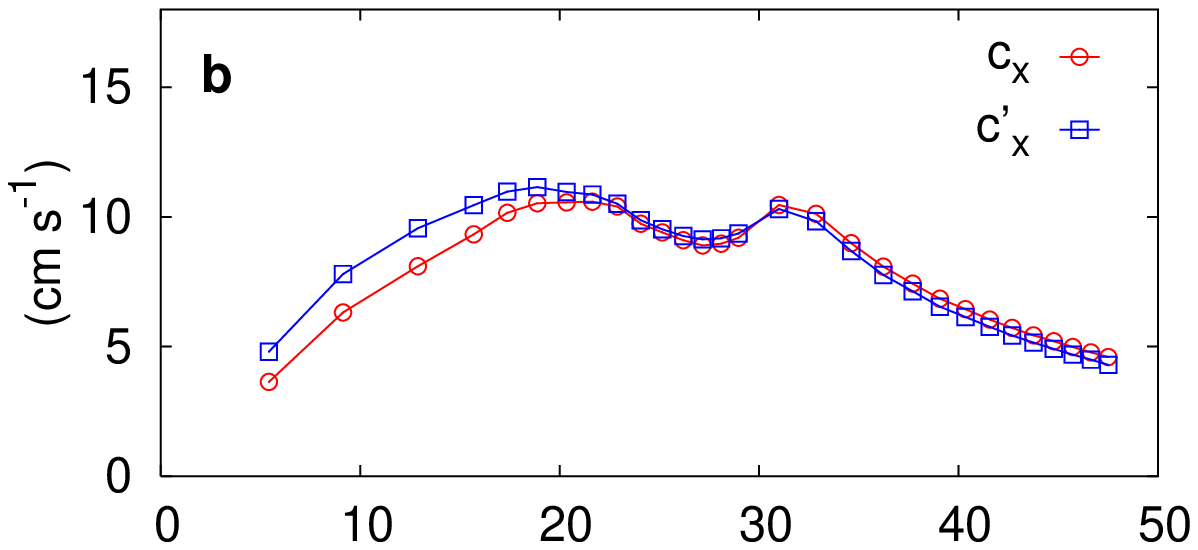}
\includegraphics[angle=0,width=3.3in]{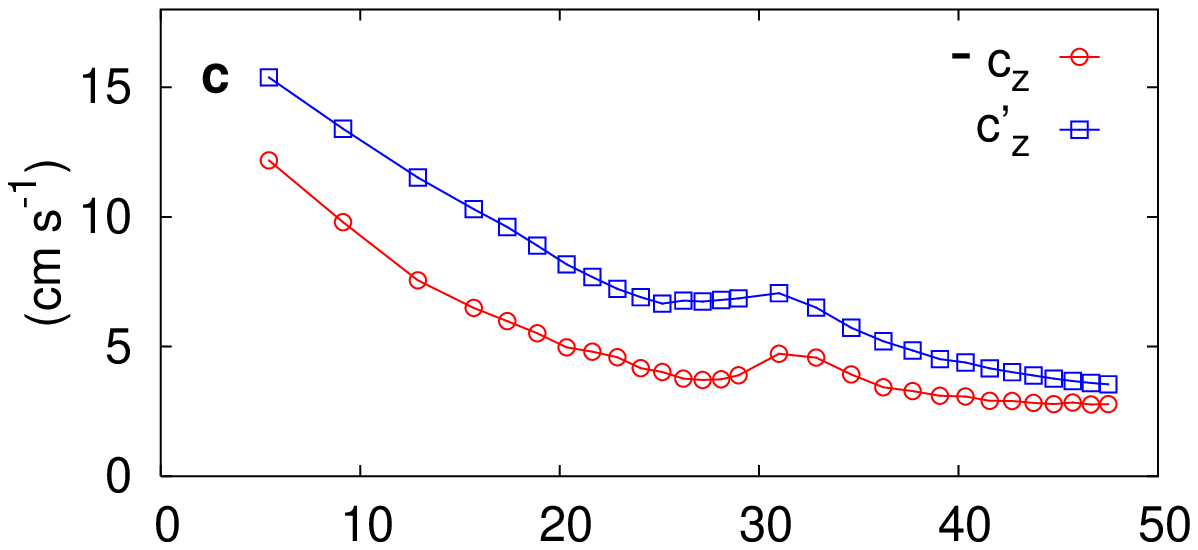}
\includegraphics[angle=0,width=3.3in]{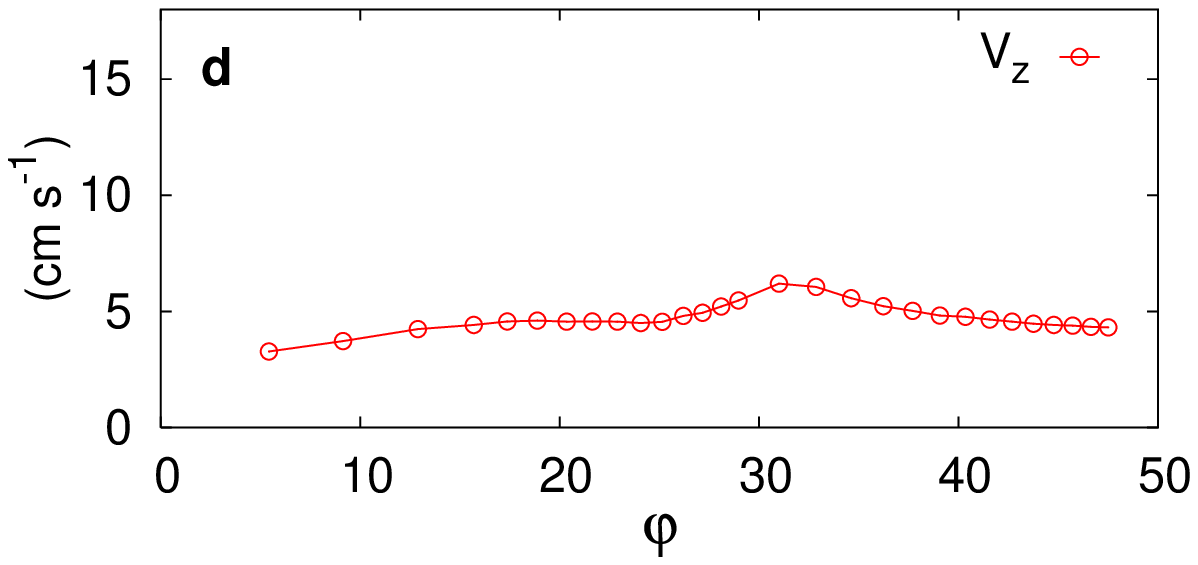}
\caption{The mean CM velocities of rods before (black) and after (red) collision
as a function of inclination angle $\phi$ in quasi-2D numerical simulations. The
parameters are $l=9.5d$, $\Gamma=3.3$, $\mu=0.3$. For comparison with experiments 
the velocities are plotted in dimensional units for experimental values of $d=0.6cm$ 
and $f=60Hz$: (a) angular velocities $\omega$ and $\omega'$; (b) vertical CM velocities $-c_z$ and $c'_z$; (c) horizontal CM velocities $c_x$ and $c'_x$; (d) the mean plate velocity at the time of collision.
}
\label{MD_velocities}
\end{center}\end{figure}

Figure \ref{MD_velocities}(a-c) the mean values of rod velocities before and
after impacts.  As seen from Fig.~\ref{MD_velocities}(a), the angular velocity before 
the collision is rather small (apparently, it decays after inelastic collisions with other rods during
flight).  Figures~\ref{MD_velocities}(b,c) shows the horizontal and vertical CM velocities of
rods just before and after collision in the laboratory frame as a
function of $\phi$. As seen from these plots, the pre- and
post-collisional velocities are close to each other. The mean velocity
of the plate at the moment of collision is shown in
Fig.~\ref{MD_velocities}(d). The velocity is only weakly dependent on
$\phi$ and is close to $V_0/2$.  In Fig.~\ref{MD_velocities}(a) one can
also see a noticeable variation of the horizontal translation velocity
near $\phi=30^o$ which however is not accompanied by a sharp drop at
$\phi\approx 20^o$ observed in 3D geometry.

\begin{figure}[h]
\begin{center}
\includegraphics[angle=0,width=3.3in]{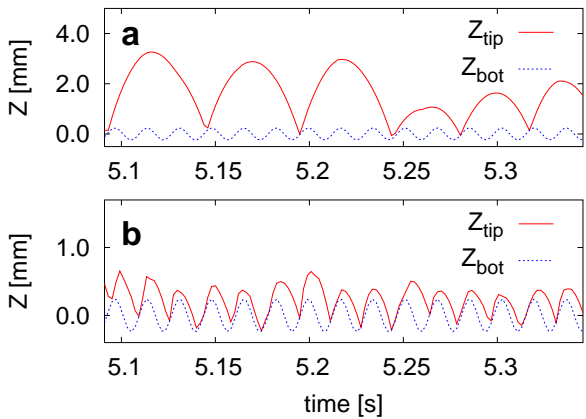}
\caption{Typical trajectories of the tip of a rod in numerical simulations for
two different tilt angles $\phi=9^o$ and 34$^o$ (other parameters are the same as in Fig.
\protect\ref{MD_velocities}
}
\label{fig:traj_phi}
\end{center}\end{figure}

We conclude that there are two different mechanisms which
independently contribute to the slowdown of rods at intermediate
values of tilt angle.  The first mechanism that only operates in 3D
geometry is related to the spontaneous formation of the non-uniformity
of rod arrangement in an annulus at intermediate inclination angles.
For small tilt, the rods are arranged in a uniform hexagonal-like
packing (see Fig.~\ref{fig:md3ddef}) with one rod near the inner wall of
the gap, next near the outer wall and so on.  This ``perfect''
arrangement may be perturbed by the presence of one or few  one-rod
defects due to geometrical frustration, however their cumulative effect
is quite small and does not change the collective motion considerably.
At $\phi\approx 18^o$ in most of the numerical experiments, the hexagonal
packing spontaneously breaks and as a result a localized region with
much larger tilt emerges.  At $\phi\approx25^o$ it involves roughly half
of the rods [Fig.~\ref{fig:md3d}(e)]. Eventually, this region spans the
whole perimeter and the dependence $c_x(\phi)$ becomes smooth again.
This defect appears to play a significant role in slowing down the rods
drift. This effect is exacerbated in numerical simulations by neglecting
the rod rotation around its axis. Therefore, rolling of rods along the
side walls is prohibited by the numerical model, and thus the sidewall
friction is effectively enhanced. 

\begin{figure}\begin{center}
\includegraphics[angle=0,width=3.0in]{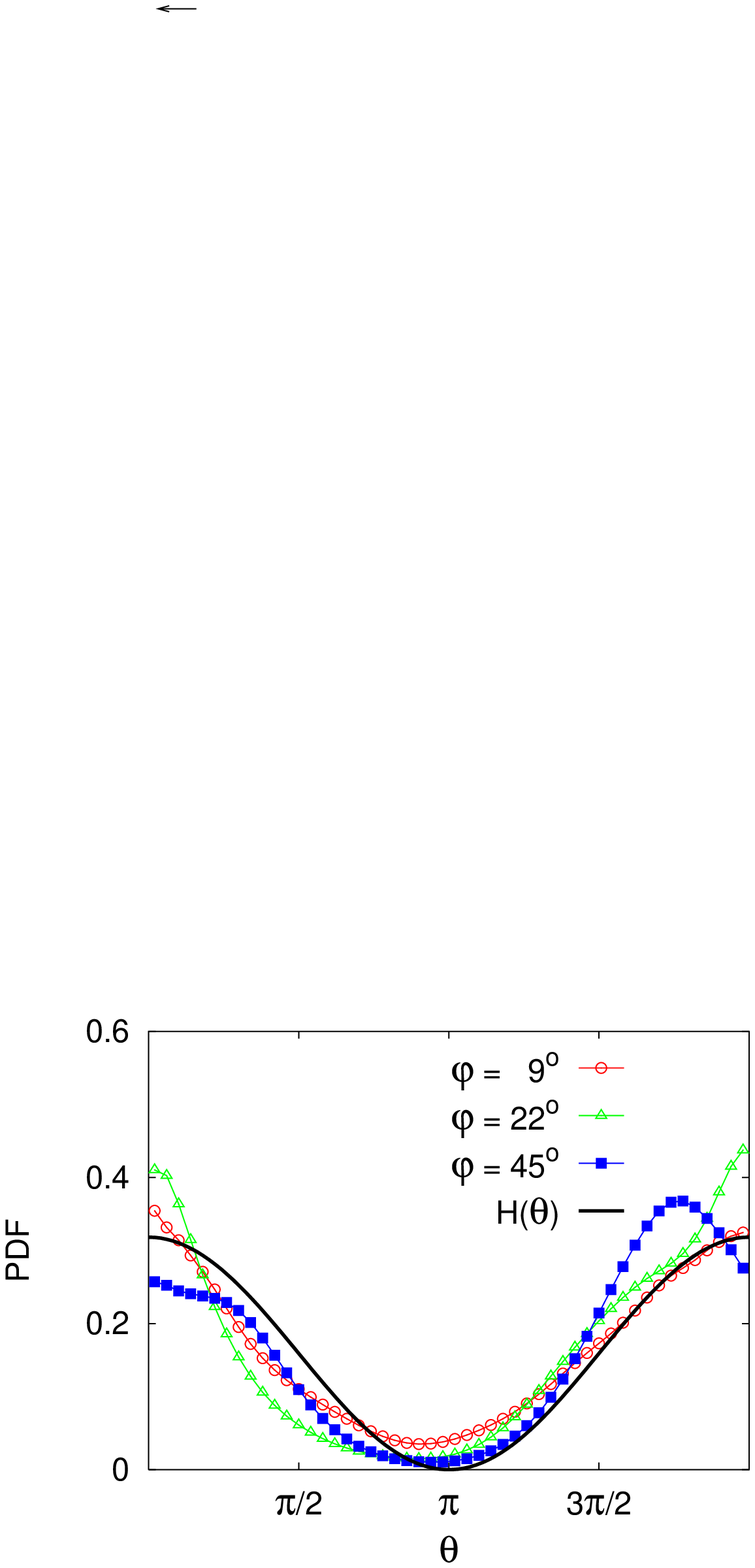}
\caption{Distribution of collision times of rods over the phases of the plate
vibration for different values of the tilt angle $\phi$. Solid line:
approximation $H(\theta)=(1+\cos\theta)/2\pi$ used for the theoretical 
analysis, where $\theta$ is the phase of the plate
velocity, $V_{pl}=V_0\cos\theta$.
Parameters are the same as in Fig.~\protect\ref{MD_velocities}.
}
\label{fig:PDF_phi}
\end{center}\end{figure}

The second mechanism operates both in 2D and 3D, and it has to do with
the bifurcation between long flights spanning more than one period of
vibrations at small tilt angles and short flights which last one period
of vibration at large tilt angles (see Figure \ref{fig:traj_phi}). This
transition occurs at $\phi\approx 30^o$ and it leads to the noticeable
difference in the distribution of the landing times over the phase of
the plate vibrations  for different tilt angles (see Figure
\ref{fig:PDF_phi}).  At large angles the landing times are mostly
confined to the phase interval in which the plate moves upward, whereas
at smaller tilt angles there is a significant probability of  collisions
during the downward motion of the plate ($\pi/2<\theta<3\pi/2$), where 
$\theta = $ mod($2\pi f t, 2\pi$).  This
bifurcation explains the bump in the dependence of the mean vertical
plate velocity at impact [see Fig.~\ref{MD_velocities}(d)] and
correspondingly the horizontal translation velocities $c_x, c_x'$ on the
tilt angle [see Fig.~\ref{MD_velocities}(b)].

Overall, our simulations show that the side walls do play an important role 
in determining the magnitude of the horizontal velocity of rods. As seen from
comparison of Figs.~\ref{fig:md3d} and \ref{fig:md2d}, the transport velocity 
in quasi-2D case is 2.5 times greater than in the annulus for the same values of parameters.
On the other hand, they allow us to develop a theoretical model of
the collective rod motion based on the observations that,
to a first approximation, the pre- and post-collision center of mass translation speeds are close 
to each other and that the angular velocity before the collision can be
neglected. 

\section{Rod collision with plate}

In this section we derive the necessary formulas for an isolated
collision between a rod and a vertically moving plate.  We confine the
analysis to the case of in-plane, eccentric, oblique frictional impact.
Our derivation is based on the classical analysis which assumes the
constant kinematic restitution coefficient (see Ref.~\cite{stronge}).

\begin{figure}\begin{center}
\includegraphics[angle=0,width=2.5in]{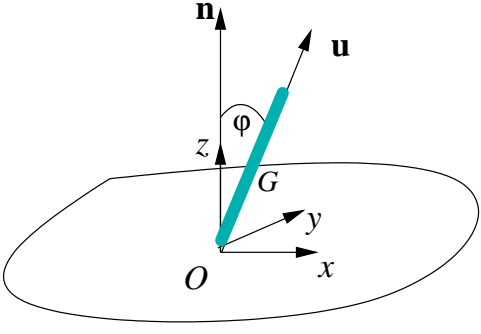}
\vskip 0.5cm
\caption{Geometry of the collision between rod and sphere.}
\label{fig_collision_RP}
\end{center}\end{figure}

Consider uniform rigid rod  of mass $m$ and length $l$ colliding with a
plate at point $O$ (see Fig~\ref{fig_collision_RP}).  We place the
system of coordinates at point $O$ so that the common normal ${\bf n}$
coincides with the ort ${\bf \hat{z}}$ and the axis of rod, prescribed
by unit vector ${\bf u}$, is in the $x-z$ plane, ${\bf u}=(\sin(\phi), 0,
\cos(\phi))$. Prior to the collision, the rod  has translational
velocity (associated with the center of mass $G$) ${\bf c} = (c_x, 0,
c_z)$, and angular velocity ${\bf \omega} = (0,\omega_y,0)$, and the
plate has only vertical velocity ${\bf V} = (0,0,V_z)$.  The
corresponding post-collisional velocities of rod are denoted by primes.

Newton's second law gives equations for translational and rotational 
motion of the rod.  In differential form they read,
\begin{eqnarray}
\label{newtont}
m d{\bf c} = d{\bf P},\\
\label{newtonr}
I d{\bf \omega} = -\frac{l}{2} {\bf u}\times d{\bf P}
\end{eqnarray}
where $I$ is the moment of inertia of a rod for its rotational
motion around the center of mass $G$. $d{\bf P}={\bf F}dt$ is the
differential of the impulse ${\bf P}$ exerted on the rod during the
collision.

The impulse acquired by the rod is the integral of the reaction force over 
the time of collision.  The reaction force depends on the relative velocity 
at the contact point (CP),
\begin{equation}
{\bf v} = {\bf c} - \frac{l}{2}{\bf \omega} \times {\bf u}  - {\bf V}.
\label{vrel}
\end{equation}

The Newton's law for the velocity of the contact point (CP) reads
\begin{equation}
m\frac{d{\bf v}}{dt} = {\bf F} + \alpha [{\bf F}-{\bf u}({\bf F}\cdot{\bf u})]
\label{dvrel1}
\end{equation}
or, in projections, 
\begin{eqnarray}
m\dot{v_x} &=& -\frac{XZ}{k^2}F_z +\frac{k^2+Z^2}{k^2}F_x 
\label{dvrelx}\\
m\dot{v_z} &=& \frac{k^2+X^2}{k^2}F_z - \frac{XZ}{k^2}F_x
\label{dvrelz}
\end{eqnarray}
where $X, Z$ are coordinates of the center of
mass (CM) of the rod, and $k$ is the radius of gyration of
the rod. In writing (\ref{dvrel1}) we assumed that the time of collision
is so short that we can neglect the changes in the plate position and
velocity. Depending on initial conditions, the impact proceeds according
to one of three different scenarios.

{\bf Slide.}
Let us denote the duration of the contact  $t_f$, so $F_z(0<t<t_f)>0,
F_z(0)=F_z(t_f)=0$.  We assume that at $t=0$, $v_x(0)=u_0>0$ and
$v_z(0)=v_0<0$.  After initial contact, the rod slides along the plate,
so $F_x=-\mu F_z$ (for brevity we dropped the subscript $rb$ of the
friction coefficient).  If 
\begin{equation}
u_* \equiv u_0-\frac{XZ+\mu (k^2+Z^2)}{mk^2}P_z(t_f)>0
\label{nostop}
\end{equation}
[here $P_z(t)=\int_0^t F_z(t)dt$] the slip in positive direction continues
throughout the contact, and $u_f\equiv u(t_f)=u_*$.
While the total impulse $P_z(t_f)$ is not known a priori,
it can be determined from the kinematic condition 
$v_f=-\epsilon v_0$ assuming that (\ref{nostop}) is
satisfied. Then integrating Eq.(\ref{dvrelz}) from $t=0$ to $t_f$ we get
\begin{equation}
P_z(t_f)=-v_0(1+\epsilon)\frac{mk^2}{k^2+X^2+\mu XZ}
\label{Pzf}
\end{equation}
and  correspondingly
\begin{equation}
P_x(t_f)=v_0(1+\epsilon)\frac{m\mu k^2}{k^2+X^2+\mu XZ}.
\end{equation}
Now we can calculate the CM velocities after the contact using
\begin{eqnarray}
c'_x&=&c_x+P_x(t_f)/m,
\label{CM_vel_x}\\
c'_z&=&c_z+P_z(t_f)/m.
\label{CM_vel_z}
\end{eqnarray}
Replacing $u_0=c_x-\omega Z $ and  $v_0=c_z-V_z+\omega X$ we get the center
of mass (CM) velocities immediately after the collision
\begin{eqnarray}
c'_x&=&c_x+\frac{\mu(1+\epsilon)k^2(c_z-V_z+\omega X)}{\mu(k^2+Z^2)+XZ},
\label{CM_vel_x1}\\
c'_z&=&cz-\frac{(c_z+\omega_x-V_z)(1+\epsilon)k^2}{\mu(k^2+Z^2)+XZ}
\label{CM_vel_z1}
\end{eqnarray}

Using (\ref{Pzf}) the no-stop sliding condition (\ref{nostop}) can be written as
\begin{equation}
u_* \equiv u_0+\frac{XZ+\mu (k^2+Z^2)}{k^2+X^2+\mu XZ}v_0(1+\epsilon)>0
\label{nostop1}
\end{equation}
If the condition (\ref{nostop1}) is violated, at a certain time $t_1$
during collision the sliding stops, $v_x(t_1)=0$. 
At $t=t_1$, the horizontal CP velocity $u_1=0$ and
the vertical CP velocity is
\begin{equation}
v_1=v_0+\frac{k^2+X^2+\mu XZ}{\mu(k^2+Z^2)+XZ}u_0.
\label{force_t1}
\end{equation}
At $t>t_1$ the contact may remain at rest or reverse the direction of
sliding. 

{\bf Slip-stick}. If 
\begin{equation}
\mu(k^2+Z^2)>XZ,
\label{slipstick}
\end{equation}
the contact sticks, and the horizontal velocity $u(t_1<t<t_f)=0$. During
this phase,
\begin{equation}
F_x=\frac{XZ}{k^2+Z^2}F_z
\label{force_stick}
\end{equation}
Assuming again the kinematic restitution law $v_f=-\epsilon v_0$, we derive the
expression for the CM velocity at the end of the collision (see
Appendix B):
\begin{eqnarray}
c'_x&=&\frac{(c_x -\omega Z)Z^2-(c_z-V_z+\omega X)(1+\epsilon)XZ}{k^2+X^2+Z^2} 
\nonumber\\
&&+ \omega Z
\label{CM_velx2_stick}\\
c'_z&=&\frac{(c_z-V_z+\omega X)[X^2-(k^2+Z^2)\epsilon]-(c_x -\omega Z)XZ}{k^2+X^2+Z^2} 
\nonumber\\
&&+V_z -\omega X
\label{CM_velz2_stick}
\end{eqnarray}
Note that for the case of slip-stick, the post-collisional velocities
are independent of the friction coefficient.

{\bf Slip reversal}. If 
\begin{equation}
\mu(k^2+Z^2)<XZ,
\label{reversal}
\end{equation}
the contact slides back after stopping. In this phase $F_x=\mu F_z$. 
Again omitting the details of derivation (see Appendix B) we give here the formulas for
the CM velocity at the end of collision: 
\begin{eqnarray}
c'_x&=&c_x +(c_x-\omega Z)\frac{2\mu k^2(k^2+X^2)}{(\mu(k^2+Z^2)+XZ)(k^2+X^2-\mu XZ)}
\nonumber\\
&&-(c_z-V_z+\omega X)\frac{\mu k^2(1+\epsilon)}{k^2+X^2-\mu XZ}
\label{CM_velx2_sliprev}\\
c'_z&=&c_z-(c_x -\omega Z)\frac{2\mu k^2 XZ}{(\mu(k^2+Z^2)+XZ)(k^2+X^2-\mu XZ)}
\nonumber\\
&&-(c_z-V_z+\omega X)\frac{k^2(1+\epsilon)}{k^2+X^2-\mu XZ}
\label{CM_velz2_sliprev}
\end{eqnarray}

\begin{figure}\begin{center}
\includegraphics[angle=0,width=2.5in]{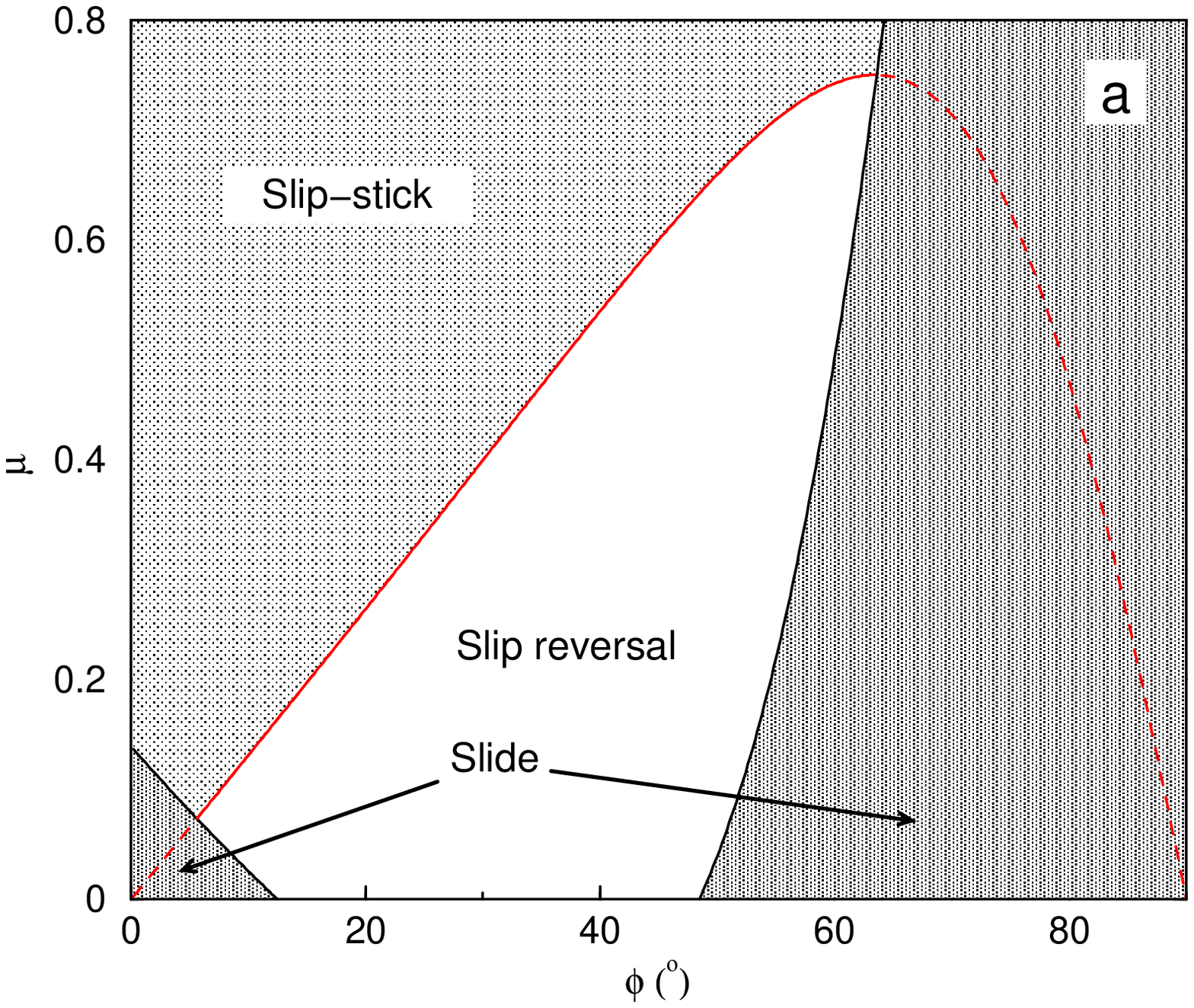}
\includegraphics[angle=0,width=2.5in]{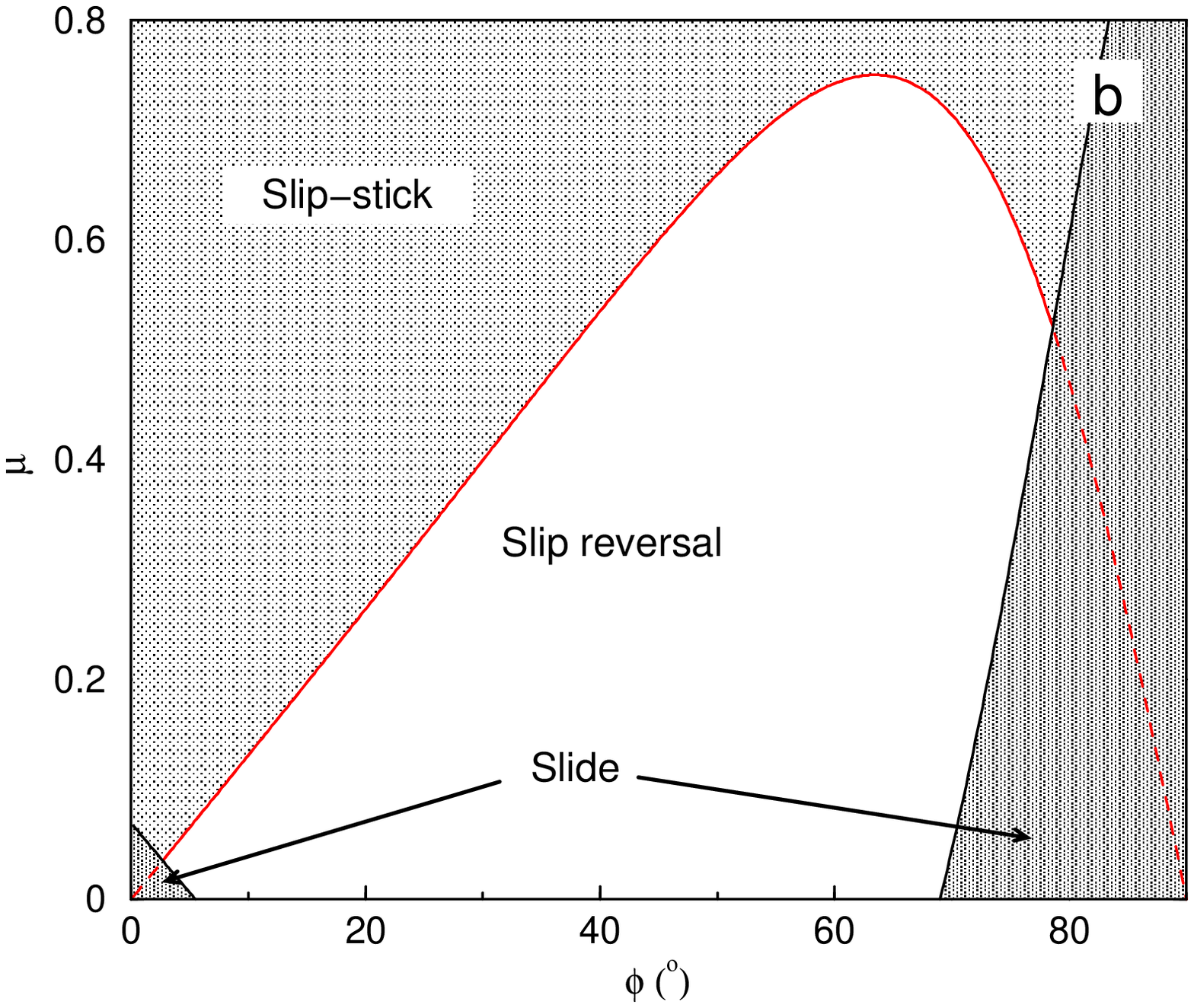}
\includegraphics[angle=0,width=2.5in]{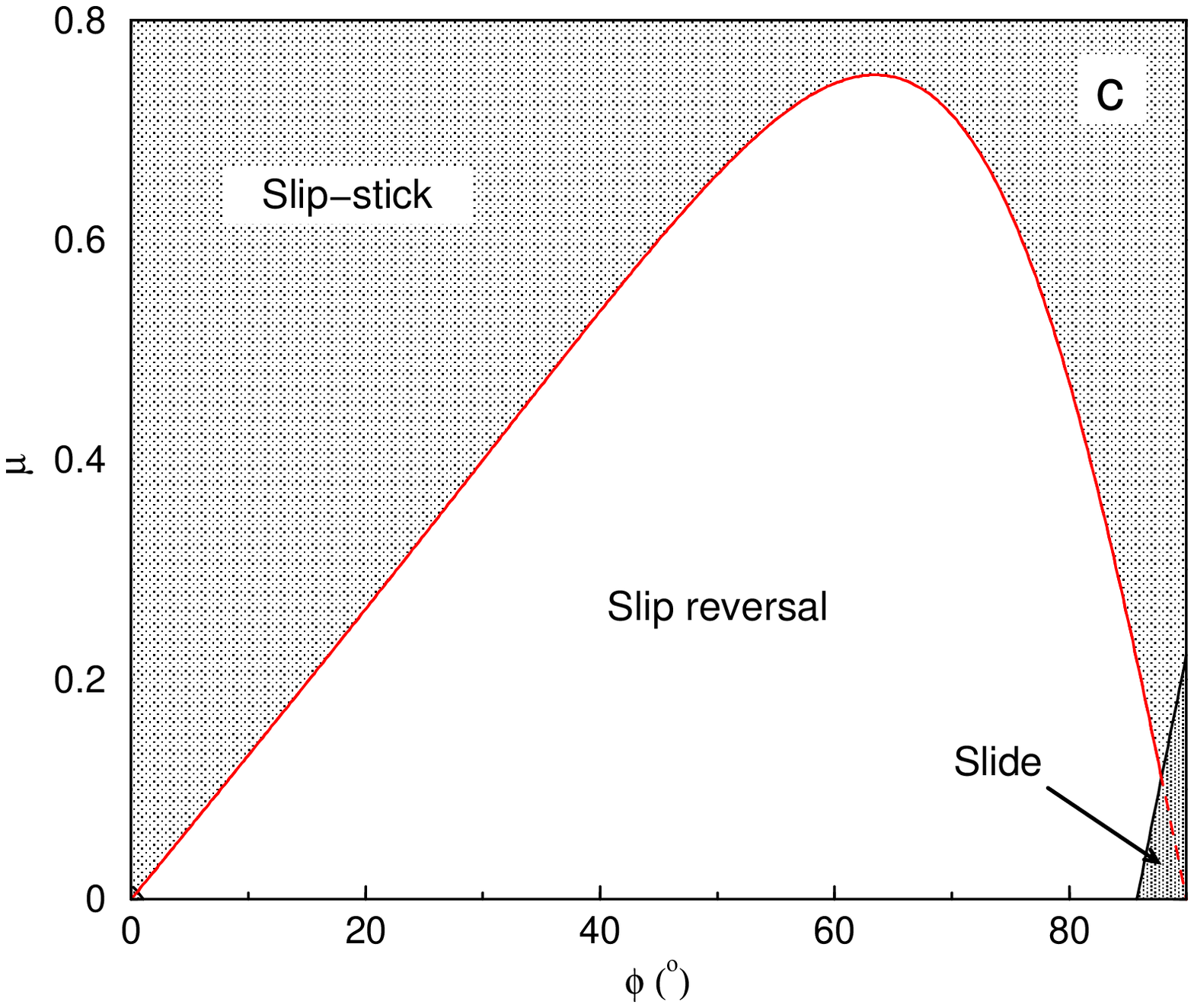}
\caption{Bifurcation diagrams of single rod collision with a plate for $\epsilon=0.8$ 
and three different values of $-c_z/c_x$: 1 (a), 2 (b), and
10 (c).}
\label{regimes}
\end{center}\end{figure}

{\bf Thin rod.}
For a thin rod of length $l$, we use values $X=l\sin\phi/2, Z=l\cos\phi/2, 
k=l/2\sqrt{3}, I=ml^2/12$. Let us first outline the boundaries of three different
regimes (continuous slide, slip=stick, and slip reversal) in the $(\phi, \mu)$ parameter 
plane. While the condition for the transition from slip-stick to slip
reversal is universal, the condition of continuous slide (\ref{nostop1}) depends on the
values of $\epsilon$ and the ratio $c_z/c_x$ and $\omega$. Figure \ref{regimes} shows
the bifurcation diagram for $\epsilon=0.8$ and three different values of $-c_z/c_x$ 
for $\omega=0$.  Typically this ratio is large, so the regime of sliding can only be
observed for small $\mu<\mu_c=-c_x[4c_z(1+\epsilon)]^{-1}$ and either
large or small $\phi$. For larger
$\mu>\mu_c$, at small $\phi$ the slip-stick regime occurs, and at larger angles
there is the slip reversal regime. The critical angle $\phi_c$ at which 
the transition from slip-stick to slip reversal occurs, is determined 
from equation for $\phi_c$ ($3\sin\phi_c\cos\phi_c=\mu(1+3\cos\phi_c^2)$.
Solving this equation, we obtain
\begin{equation}
\phi_c=\frac{1}{2}\arccos\frac{\sqrt{9-16\mu^2}-5\mu^2}{3(1+\mu^2)}
\label{phi_c}
\end{equation}
For small $\mu$, $\phi_c=\frac{4}{3}\mu +O(\mu^3)$. 
Note that the critical angle is only dependent on the friction
coefficient $\mu$ and becomes $\pi/2$ at $\mu=3/5$. At larger $\mu$, the
slip-reversal scenario does not occur at any tilt angle.

\section{Collective motion of rods}

In order to analyze the collective motion of bouncing rods using the
results for an individual rod collision obtained in the previous
section, we have to make additional assumptions regarding the
interactions of rods.  In the formulation of these assumptions,
we use the numerical and experimental observations. Referring to Figure
\ref{MD_velocities}(a-c), we assume that in the stationary translation 
regime, $\omega=0, c_x=c'_x, c_z=-c'_z$. Note that the latter simple assumption
is not very accurate for large $\Gamma$  and small
$\phi$, however using it still leads to a reasonable agreement between 
the  theory and simulations. A more accurate set of closure conditions 
would require a detailed description of the complicated interactions of
the rod during the flight between two consecutive collisions.

Adopting these simplifications, we immediately arrive to the relations
for the horizontal and vertical velocities of the rods.  Note that among
the three cases outlined above, the sliding regime cannot be realized in
the regime of stationary translation, since it would imply a continuous
decay of $c_x$. So eventually one of the two other regimes will be
established depending on the inclination angle (in a finite container,
the dynamically selected inclination angle is weakly dependent on the driving 
acceleration, see inset in Fig.~\ref{fig:md2d}(b)).

{\bf Slip-stick.}
Assuming $c_x'=c_x$, $c'_z=-c_z$, and $\omega=0$, we get 
from (\ref{CM_velx2_stick}), (\ref{CM_velz2_stick})
in the slip-stick regime ($\phi<\phi_c$)
\begin{eqnarray}
c'_x&=&\frac{2(1+\epsilon)XZ V_z}{k^2(1-\epsilon)+2X^2}
\label{cx_stick}\\
c'_z&=&\frac{(1+\epsilon) k^2V_z}{k^2(1-\epsilon)+2X^2}
\label{cz_stick}
\end{eqnarray}

For a thin rod, Eq.(\ref{cx_stick}) yields
\begin{eqnarray}
c'_x&=&\frac{6(1+\epsilon)\sin\phi\cos\phi}{1-\epsilon+6\sin^2\phi}V_z
\label{cx_stick1}
\end{eqnarray}

{\bf Slip reversal.}
In the slip-reversal regime ($\phi>\phi_c$), we 
solve Eqs.(\ref{CM_velx2_sliprev}),(\ref{CM_velz2_sliprev}) with
constraints $c_x=c'_x$, $c_z=-c'_z$, $\omega=0$. As a result, we get
\begin{eqnarray}
c'_x&=&\frac{(1+\epsilon)\mu (k^2+Z^2)V_z + (1+\epsilon)XZ V_z}
{k^2(1-\epsilon)+2X^2}
\label{cx_sliprev}\\
c'_z&=&\frac{(1+\epsilon) k^2V_z}{k^2(1-\epsilon)+2X^2}
\label{cz_sliprev}
\end{eqnarray}
Note that the vertical velocity $c'_z$ is again independent of $\mu$ and
in fact coincides with (\ref{cz_stick}).
It is easy to see that in the transition point from slip-stick to slip reversal regime
where $XZ=\mu(k^2+Z^2)$ the values of the horizontal translation speed
(\ref{cx_stick}) and (\ref{cx_sliprev}) coincide. 

For a thin rod case, we obtain from
Eq.(\ref{cx_sliprev})
\begin{eqnarray}
c'_x &=&\frac{(1+\epsilon)[\mu(1+3\cos^2\phi)+3\sin\phi\cos\phi]}{1-\epsilon+6\sin^2\phi}V_z
\label{cx_slip1}
\end{eqnarray}

\begin{figure}\begin{center}
\includegraphics[angle=0,width=3.in]{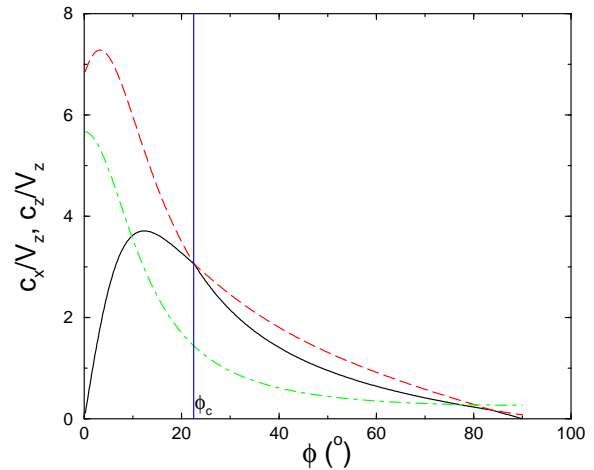}
\caption{Normalized vertical velocity after collision $c'_z/V_z$ (green
dot dashed line) and the 
translation velocity $c'_x/V_z$  (black solid line) as a function of the inclination angle 
for $\epsilon=0.65, \mu=0.3$. Red dashed shows the unphysical branches
of the slip-stick and slip reversal dependencies
(\protect\ref{CM_velx2_stick}),
(\ref{CM_velx2_sliprev}) for the horizontal post-collisional velocity. 
}
\label{cx_angle}
\end{center}\end{figure}

The vertical velocity after collision is given by
\begin{equation}
c'_z=\frac{1+\epsilon}{1-\epsilon+6\sin^2\phi}V_z
\label{cz_slip1}
\end{equation}
in both slip-stick and slip reversal regimes. Figure \ref{cx_angle}
shows the dependence of the normalized vertical velocity and the translation 
speed $c_z'/V_z, c'_x/V_z$ on the inclination angle $\phi$ for the case $\omega=0$. 
The transition from slip-stick to slip reversal dependence occurs at $\phi_c$. 

The remaining question is what is $V_z$? Obviously $V_z$ is smaller than the 
amplitude $V_0$ of the plate velocity $V=V_0\cos(\Omega t)$, because the
rods collide with the plate at different phases and not only at phase $0$
when $V=V_0$.  A simple assumption which we are going to make is that $V_z=\alpha V_0$
with a constant fitting parameter $\alpha<1$. In fact our numerical
simulations indicate that $\alpha$ is close to 0.5 but varies slightly with
$\phi$ because the landing phase distribution depends 
on $\phi$ (see Figure \ref{fig:PDF_phi}), but for the sake of simplicity  we shall 
ignore this dependence. The value $\alpha=0.5$ is obtained if we approximate the 
distribution of collision phases as
$H(\theta)=(1+\cos\theta)/2\pi$, where $\theta$ is the phase of the plate
velocity, $V_{pl}=V_0\cos\theta$. Then we obtain for the average plate speed at
collision,
\begin{equation}
V_z=V_0(2\pi)^{-1}\int_0^{2\pi} \cos\theta (1+\cos\theta)d\theta=V_0/2
\label{Vz}
\end{equation}

At small $0<\Gamma-1< 1$, the distribution deviates significantly from
$H(\theta)$. The rods only leave the plate for short flights near
the top position of the plate at which the vertical acceleration is
smaller than $-g$, and the vertical velocity $V$ is close to zero. Due
to inelasticity, after landing the rod may perform a few more smaller
bounces before coming to rest until the next period. While it is
difficult to describe this regime analytically, one can expect that
$V_z\propto \Gamma -1$ at small $\Gamma$. 

\section{Discussion}

\begin{figure}\begin{center}
\includegraphics[angle=0,width=3.in]{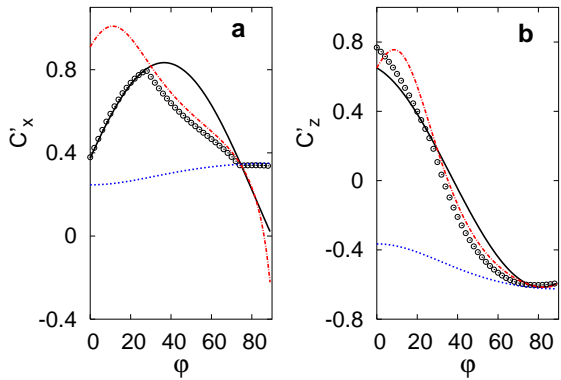} 
\caption{Post-collisional CM velocities $c'_x$ and $c'_z$ as a function of tilt
angle for $\mu=0.4$ and $-c_z/c_x=2$.  Symbols indicate the results of numerical
 simulations, and lines show theoretical predictions for different collision scenarios
using constant coefficient of restitution, $\epsilon=0.65$. Solid lines denote slip-stick,
dash lines denote continuous slip, and  dash-dot lines denote slip-reversal.
}
\label{singlemd}
\end{center}\end{figure}

In this section we compare the theoretical results with 
numerical simulations for the quasi-2D case.
First, we tested the theoretical predictions for the isolated rod
bounced off the plate, and found a good agreement between formulas
(\ref{CM_vel_x1}), (\ref{CM_vel_z1}), 
(\ref{CM_velx2_stick}), (\ref{CM_velz2_stick}), 
(\ref{CM_velx2_sliprev}), (\ref{CM_velz2_sliprev}), 
and numerics (see Fig.~\ref{singlemd}).  One can clearly see the
transitions between three different regimes of rods bouncing:
slide, slip-stick, and slip reversal.  A slight difference between
the theory and simulations in the vertical velocity at small tilt angles
is related to the above-mentioned variations of the kinematic restitution
coefficient with tilt angle in soft-particle MD simulations which was
ignored in an analytical calculations. 

\begin{figure}\begin{center}
\includegraphics[angle=0,width=3.0in]{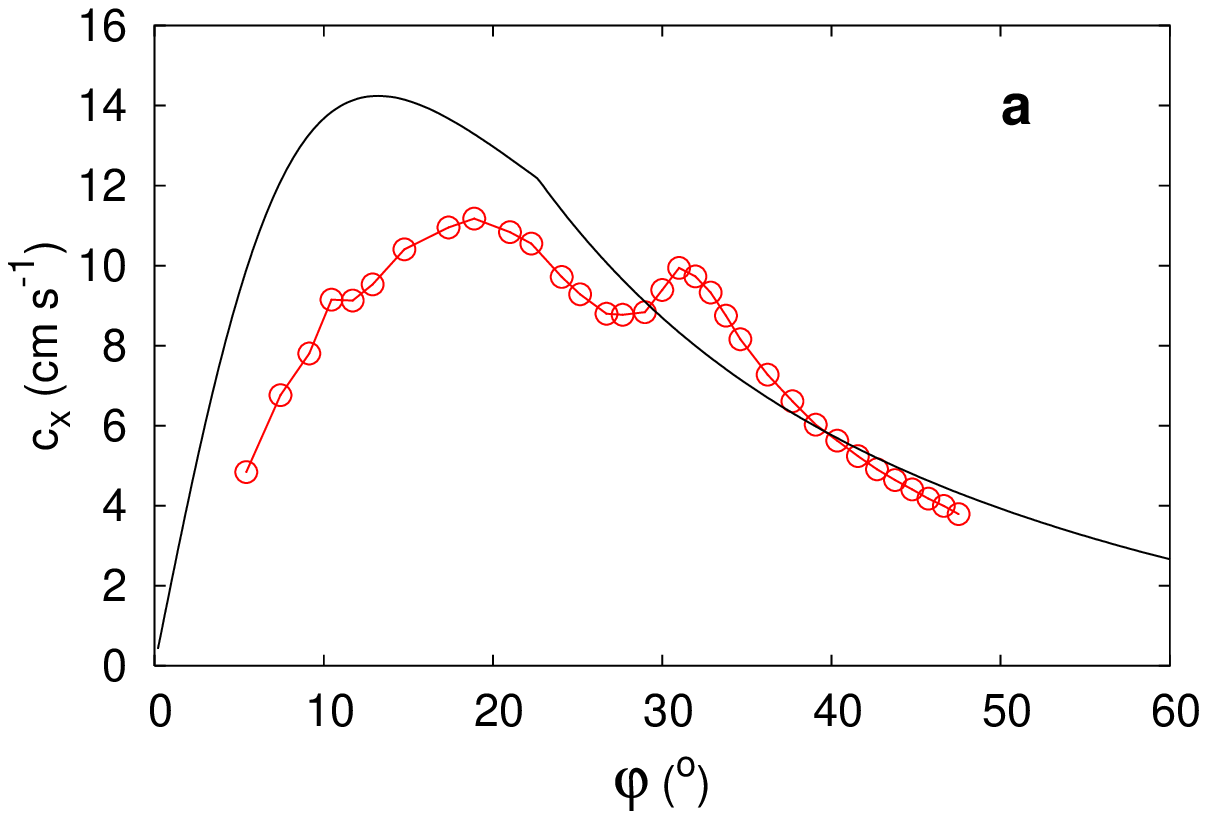}
\includegraphics[angle=0,width=3.0in]{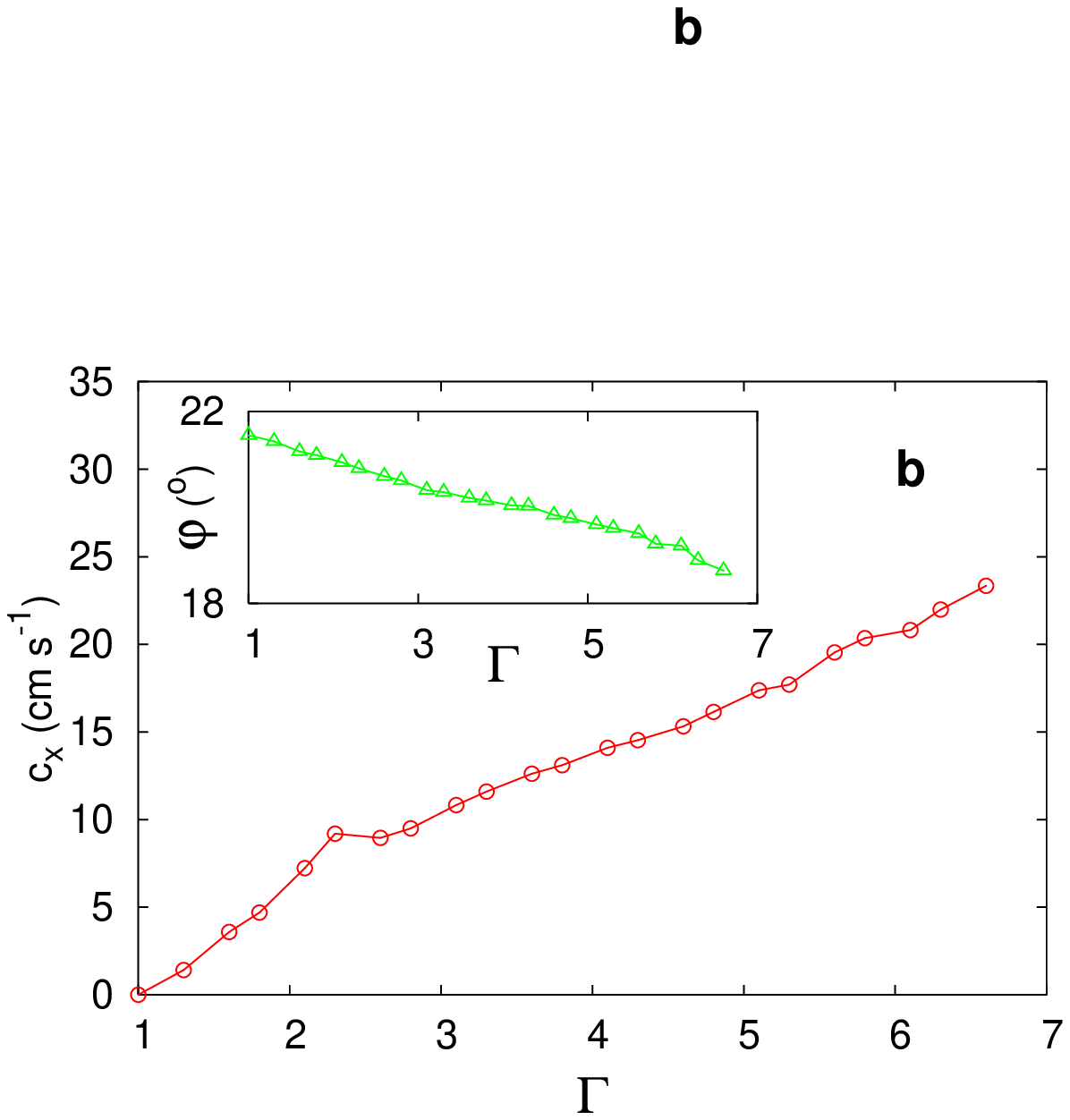}
\includegraphics[angle=0,width=3.3in]{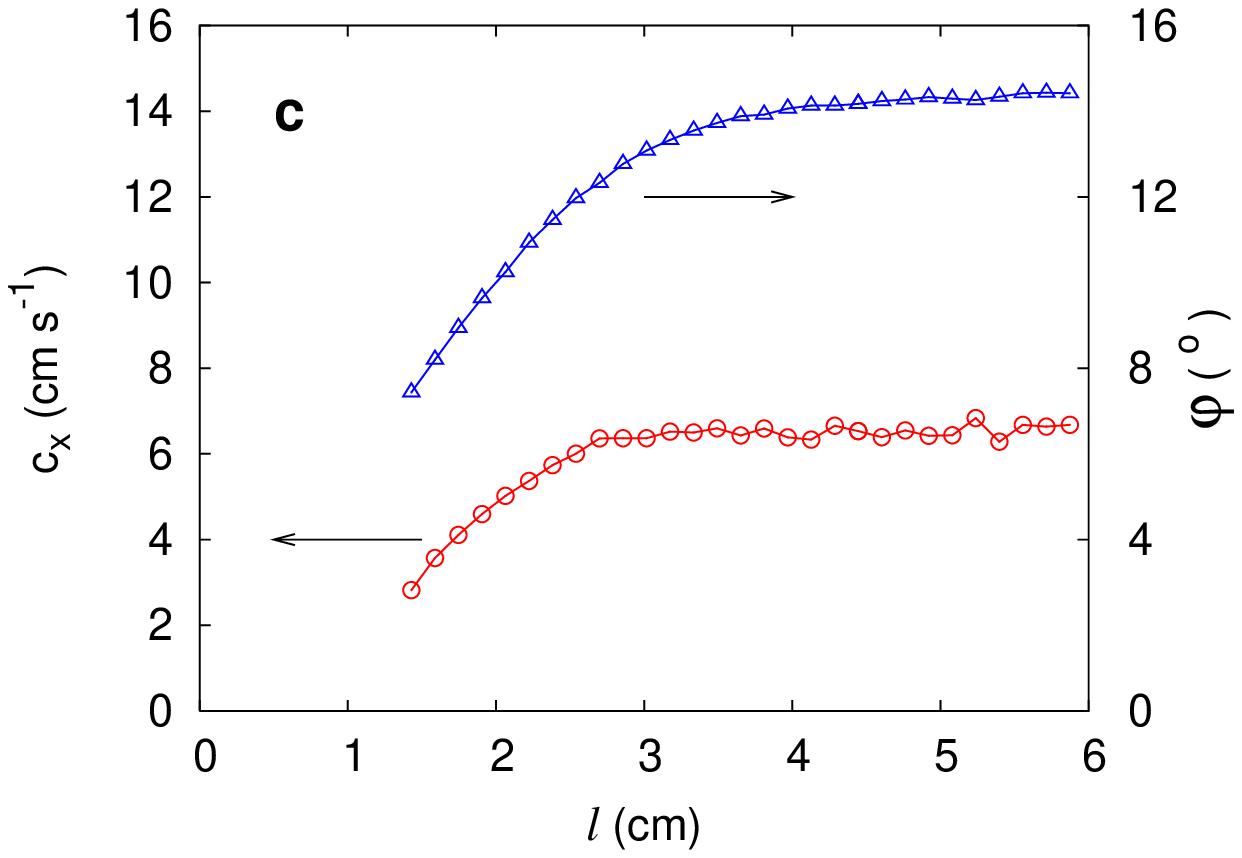} 
\caption{Results of simulations in quasi-2D geometry. $N=40$ rods are placed in a periodic domain of different lengths $L$ which determine the tilt angle $\phi_0$. ({\em a}) -  average horizontal velocity  of rods as a 
function of the inclination angle for $\Gamma=3.3$, solid line -  theory (\ref{cx_stick}), (\ref{cx_slip1}), 
(\ref{Vz})  with $\epsilon=0.65, \mu=0.3$;
({\em b}) - average horizontal velocity  of rods and average tilt (inset) as a function $\Gamma$ for $L=43$;
({\em c}) average horizontal velocity of rods and average tilt 
as a function of the length of rods $l=h_0+1$,
where $h_0$ is the distance between the centers of spherical caps.
$N=40$ rods are placed in a periodic domain of length $L=41.4$ \cite{note:cxL}, the bottom is
oscillating at frequency $f=60 Hz$ and acceleration is $\Gamma=2.5$.
}
\label{fig:md2d}
\end{center}
\end{figure}

A comparison between the theory and MD simulations for the mean translation
velocity is shown in Figure \ref{fig:md2d}.  Figure \ref{fig:md2d}(a)
shows the horizontal translation velocity as a function of the mean tilt
angle. The overall dependence is in reasonable 
agreement with the theory, however some noticeable
differences are obvious. This should not be surprising, given the crude
assumptions made to describe the dynamics of rods during flights between
collisions. As mentioned above, the `bump' visible at $\phi\approx 30^o$
is related to a transition from the regime of `long' flights that span
more than one period of vibrations,  to the `short' flights that last a
fraction of the period of vibrations.  As seen in Figure
\ref{fig:PDF_phi}, these two regimes are characterized by significantly
different distributions of the collision times over the vibration
phases. 

Figure \ref{fig:md2d}(b) shows the $\Gamma$ dependence of the  horizontal
translation velocity. As expected from the theory, and seen in
experiments, the horizontal translation velocity is linearly
proportional to $\Gamma$ at large $\Gamma$. Unlike the annular case, the
translation velocity turns zero at $\Gamma=1$, and indeed it
grows as $\Gamma-1$ at small $\Gamma-1$.  Figure \ref{fig:md2d}(c)
addresses the question of the rod length dependence of the horizontal
translation velocity. According to the theory for thin rods, $c_x$  should be
independent of $l$. On the other hand, the drift should
disappear when the aspect ratio of the rod approaches 1 (the case of
spherical particles). As seen in Figure \ref{fig:md2d}(c), the
translation velocity grows linearly at small $l>1$, but this growth
saturates at $l\approx 3$ after which the translation velocity is
independent of $l$ in agreement with the theory.

There are several possible sources of discrepancies between the theory
and numerics (and experiments). First, in describing individual
collisions we made an implicit assumption that the collision time is
small as compared with the period of oscillations. This assumption may
break for high frequency vibrations or in the regime of small $\Gamma$
when rods spend a significant portion of the period in contact with the
plate.  Furthermore, we used the simplest closure assumptions to relate
the horizontal and vertical velocities of rods after and before the
flight. While the relation $c_x=c_x'$ appears to work well throughout
the range of parameters corresponding to experimental conditions, the
other condition $c_z=-c_z'$ holds only approximately. Our numerical
experiments showed a significant (up to 50\%) deviations from this
simple relation at large $\Gamma$, when many inelastic collisions occur
during the flight. We were unable to describe this effect theoretically,
and chose to sacrifice the accuracy of comparison rather than introduce
an unknown fitting parameter $-c_z/c_z'$. 

Comparison between 2D and 3D simulations (Figs. \ref{fig:md3d},
\ref{fig:md2d}) shows that the characteristic translation velocity in 3D
case is 2.5 times smaller than in 2D case with the same material
parameters. This difference  may be accounted for by the
frictional interaction with side walls.  We systematically studied the
dependence of the translation velocity in the annulus on the friction
coefficient $\mu_{rw}$ between the rods and the side walls, and found
that indeed it varies strongly with $\mu_{rw}$, in particular, the
translation velocity at $\mu_{rw}=0.3$ is 2.5 times smaller than $c_x$ at
$\mu_{rw}=0$ (see Fig. \ref{fig:md2d}(b)).  We also analyzed the
dependence of the translation velocity on the friction coefficient with
the bottom, and found that for large $\mu_{rb}$ the translation velocity
becomes independent of $\mu_{rb}$. This is consistent with the
theoretical argument that at large $\mu_{rb}$ the slip-stick scenario occurs
for an arbitrary tilt angle $\phi$. 
 
As a conclusion, we studied experimentally and theoretically the drift
of anisotropic particles (rods) on a vibrated plate. The experiments in
the annulus showed the robust drift of rods in the direction of their
tilt, at the normalized vertical acceleration $\Gamma>1.5$. For smaller
values of $1<\Gamma<1.5$, very small {\em reverse} drift was observed.
We developed a novel numerical algorithm which allowed us to study the
interaction of rods in soft-particle MD simulations. Simulations of rods
in an annulus with parameter closely matched experiment, revealed very
similar behavior, both qualitatively and quantitatively. 

Out theoretical description of the rod translation is based on the
detailed analysis of frictional collisions between an individual rod and
the moving plate. The effects of collective interaction of rods during
flights between collisions are taken into account using the simplest
phenomenological closure conditions based on the experimental findings
and MD simulations. As a result, closed formulas for the horizontal
translation velocity are obtained.  A direct comparison between the
theory and experiments is complicated by the role of interaction of rods
with frictional side walls which is unaccounted for in the theory.
However, we found a reasonable agreement between the theory and numerics
for quasi-2d geometry when rods are confined to the $x-z$ plane with
periodic in $x$ boundary conditions. Since the same numerical code
describes well the experiment in the annulus geometry, we infer that the
theory correctly captures the mechanism of the rod translation in
experiment. 

Some more subtle effects are, however, are difficult to model
theoretically. The (very slow) reverse drift of rods for small $\Gamma$
is presumably due to the small negative value of average $V_z$, however
to calculate the mean $V_z$ one needs a detailed knowledge of the
distribution of landing times for small $\Gamma$ which is difficult to
obtain theoretically. 

We thank Daniel Blair for help with performing preliminary experiments
and for many helpful discussions. This work was funded by the NSF Grant \#
DMR-9983659 (AK) and the U.S. Department of Energy, Office of Basic Energy
Sciences, under grant \# DE-FG03-95ER14516 (DV and LT).

\appendix
\section{Molecular Dynamics Algorithm}

In our MD algorithm, two virtual spheres of diameter $d$,
with centers at ${\bf r}_i$ and ${\bf r}_j$, and with velocities ${\bf
v}_i$
and ${\bf v}_j$, interact via normal and tangential forces,
${\bf F}_{ij} = F_n {\bf n}_{ij} +{\bf F}_t$, $F_n=k_n \delta^{3/2} - \gamma_n M_e \delta  v_n$.
We introduce tangential spring with deflection obtained by the integration of
tangential velocity through the period of impact, ${\bf s}=\int d\tau{\bf v}_t$, then
the tangential force component is defined separately in two cases:
${\bf F}_t = - k_t \delta{\bf s} - \gamma_t M_e \delta  v_t $, 
for stick phase, and ${\bf F}_t =- \mu_{rr}F_n{\bf t}_{ij}$ for slip phase. During the slip
phase the magnitude of ${\bf s}$ is adjusted to fulfill $|{\bf F}_t|=\mu_{rr}|F_n|$. 
Here $M_e=M/2$ is reduced mass for rod-rod collision,  $m$ is the mass of the rod,
$\delta=d-r_{ij}$ and $v_n = {\bf v}_{ij}\cdot{\bf n}_{ij}$ are the
overlap and relative velocity in the direction of normal, ${\bf n}_{ij}=({\bf r}_i-{\bf
r}_j)/r_{ij}$, while tangential direction ${\bf t}_{ij}={\bf v}_t/v_t$ is specified by
the relative tangential velocity ${\bf v}_t = {\bf v}_{ij}- v_n{\bf n}_{ij}$, 
$\mu_{rr}$ is coefficient of friction between rods.  MD is performed in reduced units, all quantities are normalized
by an appropriate combination of the diameter $d$, mass of virtual sphere $m$,
and gravitation acceleration $g$. 
Typical values of material parameters are, $k_n=5\ 10^6 (mg/D)$,$k_t=95k_n$,
$\gamma_n=\gamma_t=4\ 10^2 (g/D)^{1/2}$. The coefficients of friction for rod-rod and rod-bottom collisions are
$\mu_{rr}=0.3$,$\mu_{rb}=0.3$ respectively. Unless specified otherwise for interactions with walls we also used 
$\mu_{rw}=0.3$.

To expedite the integration of Newton's equation we used simple leapfrog algorithm \cite{allen87} 
with constant time step $\Delta t = 2.0\ 10^{-5} (d/g)^{1/2}$, however we tested that
application of more accurate integration scheme such as $5^{th}$-order Gear predictor-corrector
did not introduce considerable changes.

Our choice of the values of material parameters is neither optimal for
the comparison with experimental data nor unique. Because we observed very good agreement with
the theoretical description for a single collision (see Fig. \ref{singlemd})
we expect our algorithm to capture details of {\em short-term}  collision with plate.
However, for a long-term collision it may not be accurate.

\section{Derivation of reflection coefficients}

{\bf Slip-stick.}
The stopping condition $u(t_1)=0$ gives the total vertical impulse exerted 
by a plate on a rod for $0<t<t_1$,
\begin{equation}
P_z(t_1)=u_0\frac{mk^2}{\mu(k^2+Z^2)+XZ}
\label{Pz_t1}
\end{equation}
and correspondingly
\begin{equation}
P_x(t_1)=-u_0\frac{m\mu k^2}{\mu(k^2+Z^2)+XZ}
\label{Px_t1}
\end{equation}

Vertical velocity at the end of the contact 
\begin{eqnarray}
v_f&=&v_1+\frac{k^2+X^2+Z^2}{k^2+Z^2}\int_{t_1}^{t_f}F_z=
v_0+\label{vf}\\
&+& \frac{k^2+X^2+\mu XZ}{\mu(k^2+Z^2)+XZ}u_0+
\frac{k^2+X^2+Z^2}{m(k^2+Z^2)}(P_z(t_f)-P_z(t_1))\nonumber
\end{eqnarray}
Assuming the kinematic restitution law $v_f=-\epsilon v_0$ and using
(\ref{Pz_t1}),(\ref{Px_t1}), we get
\begin{eqnarray}
&&P_z(t_f)=u_0\frac{mk^2}{\mu(k^2+Z^2)+XZ}
\label{Pz_stick_f}\\
&-& \left[v_0(1+\epsilon)+\frac{k^2+X^2+\mu
XZ}{\mu(k^2+Z^2)+XZ}u_0\right]\frac{m(k^2+Z^2)}{k^2+X^2+Z^2}
\nonumber
\end{eqnarray}
\begin{eqnarray}
&&P_x(t_f)=-u_0\frac{m\mu k^2}{\mu(k^2+Z^2)+XZ}
\label{Px_stick_f}\\
&-&\left[v_0(1+\epsilon)+\frac{k^2+X^2+\mu
XZ}{\mu(k^2+Z^2)+XZ}u_0\right]\frac{mXZ}{(k^2+X^2+Z^2)}
\nonumber
\end{eqnarray}
Now we can calculate the CM velocities after the contact.
Substituting (\ref{Pz_stick_f}),(\ref{Px_stick_f}) with
$u_0=c_x -\omega Z $ and  $v_0=c_z-V_z+\omega X$ into
(\ref{CM_vel_x}),(\ref{CM_vel_z}), we get
(\ref{CM_velx2_stick}),(\ref{CM_velz2_stick}).

{\bf Slip reversal.}
Using (\ref{dvrelx}),(\ref{dvrelz} and the slip condition $F_x=\mu F_z$ for $t_1<t<t_f$, 
we obtain the horizontal and vertical velocities at $t_f$,
\begin{eqnarray}
u_f&=&\frac{\mu(k^2+Z^2)-XZ}{mk^2}(P_z(t_f)-P_z(t_1))
\label{uf_sliprev}\\
v_f&=&v_1+\frac{k^2+X^2-\mu XZ}{mk^2}(P_z(t_f)-P_z(t_1))
\label{vf_sliprev}
\end{eqnarray}
Again assuming kinematic restitution condition $v_f=-\epsilon v_0$ and
using (\ref{Pz_t1}),(\ref{Px_t1}), we get for the
impulse during the reversal phase
\begin{eqnarray}
&&P_z(t_f)=u_0\frac{mk^2}{\mu(k^2+Z^2)+XZ}
\label{impulse_reversal_z}\\
&-&\frac{mk^2}{k^2+X^2-\mu XZ}
\left[(1+\epsilon)v_0+\frac{k^2+X^2+\mu XZ}{\mu(k^2+Z^2)+XZ}u_0\right]
\nonumber\\
&&P_x(t_f)= -u_0\frac{\mu mk^2}{\mu(k^2+Z^2)+XZ}
\label{impulse_reversal_x}\\
&-&\frac{\mu mk^2}{k^2+X^2-\mu XZ}
\left[(1+\epsilon)v_0+\frac{k^2+X^2+\mu XZ}{\mu(k^2+Z^2)+XZ}u_0\right]
\nonumber
\end{eqnarray}

The final horizontal velocity of CP
\begin{equation}
u_f= -\frac{\mu(k^2+Z^2)-XZ}{k^2+X^2-\mu XZ}
\left[(1+\epsilon)v_0+\frac{k^2+X^2+\mu XZ}{\mu(k^2+Z^2)+XZ}u_0\right]
\label{uf_reversal}
\end{equation}

Substituting (\ref{impulse_reversal_z}),(\ref{impulse_reversal_x}) into
(\ref{CM_vel_x})-(\ref{CM_vel_z}), we get  the CM velocities after the
contact for the case of slip reversal
\begin{eqnarray}
&&c'_x=u_0\left[1-\frac{\mu k^2}{\mu(k^2+Z^2)+XZ}\right]
\label{CM_velx1_sliprev}+\omega Z\\
&-&\frac{\mu
k^2}{k^2+X^2-\mu XZ}
\left[(1+\epsilon)v_0+\frac{k^2+X^2+\mu XZ}{\mu(k^2+Z^2)+XZ}u_0\right],
\nonumber\\
&&c'_z=v_0+u_0\frac{k^2}{\mu(k^2+Z^2)+XZ}
\label{CM_velz1_sliprev}\\
&-&\frac{k^2}{k^2+X^2-\mu XZ}
\left[(1+\epsilon)v_0+\frac{k^2+X^2+\mu XZ}{\mu(k^2+Z^2)+XZ}u_0\right]
+V_z-\omega X.
\nonumber
\end{eqnarray}
Substituting $u_0=c_x -\omega Z $,  $v_0=c_z-V_z+\omega X$, we get
(\ref{CM_velx2_sliprev}),(\ref{CM_velz2_sliprev}).



\begin{thebibliography}{11}
\bibitem{note1} With this approach the steric exclusion is modeled more accurately
than in the case when the rod is approximated by a row of fixed spheres. On the
other hand the most time-taking operation of soft-spheres MD, finding the overlap,
is reduced to minimization of the quadratic form (finding the 
closest point between two segments).
\bibitem{note:cxL} In contrast with the rest of the paper we assumed that rods are actual spherocylinders
with dimensionless mass $1+3/2 h_0$ and moment of inertia $1/10+15/32 h_0 + h_0^2 + 1/8 h_0^3$
\end{thebibliography}
\end{document}